\documentclass[11pt]{article}

\usepackage[a4paper,margin=1in]{geometry}

\usepackage[T1]{fontenc}
\usepackage[utf8]{inputenc}
\usepackage{lmodern}
\usepackage{microtype}

\usepackage{amsmath,amssymb,mathtools,bm}

\usepackage{graphicx}
\usepackage{xcolor}

\usepackage{booktabs}
\usepackage{multirow}
\usepackage{array}

\usepackage{enumitem}

\usepackage{caption}
\usepackage{subcaption}

\usepackage{listings}
\definecolor{codekw}{RGB}{0,0,170}
\definecolor{codestr}{RGB}{163,21,21}
\definecolor{codecmt}{RGB}{0,120,0}
\definecolor{codebg}{RGB}{248,248,248}
\lstdefinestyle{pps}{
  language=Python,
  basicstyle=\ttfamily\footnotesize,
  keywordstyle=\color{codekw}\bfseries,
  stringstyle=\color{codestr},
  commentstyle=\color{codecmt}\itshape,
  backgroundcolor=\color{codebg},
  frame=single,
  rulecolor=\color{black!20},
  showstringspaces=false,
  breaklines=true,
  columns=fullflexible,
  numbers=none,
  xleftmargin=0.6em,
  aboveskip=0.9em,
  belowskip=0.7em
}
\usepackage{hyperref}
\usepackage[nameinlink]{cleveref}

\crefname{figure}{Fig.}{Figs.}
\Crefname{figure}{Fig.}{Figs.}
\crefname{table}{Table}{Tables}
\Crefname{table}{Table}{Tables}
\crefname{section}{Sec.}{Secs.}
\Crefname{section}{Sec.}{Secs.}
\crefname{subsection}{Sec.}{Secs.}
\Crefname{subsection}{Sec.}{Secs.}
\crefname{subsubsection}{Sec.}{Secs.}
\Crefname{subsubsection}{Sec.}{Secs.}
\crefname{equation}{Eq.}{Eqs.}
\Crefname{equation}{Eq.}{Eqs.}
\crefname{lstlisting}{Listing}{Listings}
\Crefname{lstlisting}{Listing}{Listings}
\crefname{appendix}{Appendix}{Appendices}
\Crefname{appendix}{Appendix}{Appendices}

\usepackage{indentfirst}
\usepackage{orcidlink}

\setlist[itemize]{leftmargin=1.5em}
\setlist[enumerate]{leftmargin=1.8em}

\hypersetup{
    pdftitle={PADO-Pauli: A GPU-Accelerated Implementation of Pauli Propagation},
    pdfauthor={Hyunwoo Kim, Youngseok Lee},
    colorlinks=true,
    linkcolor=blue!60!black,
    citecolor=blue!60!black,
    urlcolor=blue!60!black
}

\DeclareRobustCommand{\pado}{\texorpdfstring{{\normalfont\textsf{PADO-Pauli}}}{PADO-Pauli}}
\DeclareRobustCommand{\pps}{\texorpdfstring{{\normalfont\textsc{PPS}}}{PPS}}
\DeclareRobustCommand{\qaoa}{\texorpdfstring{{\normalfont\textsc{QAOA}}}{QAOA}}
\newcommand{\ket}[1]{\lvert #1 \rangle}
\newcommand{\bra}[1]{\langle #1 \rvert}
\DeclareMathOperator{\Tr}{Tr}

\title{\textbf{\textsf{PADO-Pauli}: A GPU-Accelerated Implementation of Pauli Propagation}}

\author{
    \begin{tabular}[t]{@{}c@{\hspace{3em}}c@{}}
        Hyunwoo Kim\,\orcidlink{0009-0009-7190-4940} & Youngseok Lee\,\orcidlink{0009-0008-7970-1921} \\[0.2em]
        \texttt{hw\_kim@norma.co.kr} & \texttt{ys\_lee@norma.co.kr}
    \end{tabular}
}

\date{Quantum AI Team, NORMA, Inc., Seoul, Republic of Korea}

\begin{document}

\maketitle

\begin{abstract}
Variational quantum algorithms utilize parameterized circuits many times during training, at new parameter values, over batches of data, and with gradients. Pauli propagation approximates the expectation values of such circuits classically by evolving the observable in the Heisenberg picture as a truncated sum of Pauli strings. We present \pado, a GPU engine for Pauli propagation that compiles a circuit and its observables once into sparse matrices in which each rotation angle contributes only a cosine and a sine factor, so every later evaluation reuses the compiled program. It evaluates batches of parameters and data inputs embedded as rotation angles. By default its gradients come from a vector-Jacobian product registered with PyTorch's automatic differentiation, so they flow through the angles to any classical layer that produces them. \pado\ also provides truncation by Pauli weight and coefficient size, and noise channels. Zero-filtering removes terms that do not contribute to an expectation value in the all-zero reference state and makes the batched forward pass about 1.5 times faster. Compared to cuPauliProp on an NVIDIA A100, \pado\ evaluates up to 10 times faster on 4 to 12 qubits and uses up to 3.1 times less memory on 16 qubits. We check its accuracy against exact statevector simulation up to 28 qubits. These features make \pado\ a practical toolkit for training variational quantum algorithms, including quantum machine learning models.
\end{abstract}

\clearpage

\setcounter{tocdepth}{3}
\tableofcontents

\clearpage

\section{Introduction}

Variational quantum algorithms and quantum machine learning models typically evaluate a parameterized circuit, update its parameters with a classical optimizer, and repeat. The same circuit structure is therefore queried many times for expectation values of Hamiltonians or cost operators written as sums of Pauli strings.

Classical simulation of this loop has been dominated by Schr\"odinger picture methods, in particular full statevector simulation, which is exact and general but tracks the whole state, costing memory and runtime exponential in the system size. An alternative is to work in the Heisenberg picture \cite{Gottesman1998TheHR} and propagate observables through the circuit in reverse gate order. This idea has grown into a family of methods now called \emph{Pauli propagation} \cite{PhysRevA.99.062337}.

The method has been reported to be competitive with leading simulators for certain tasks \cite{6vd7-l9bn}. Pauli propagation, sparse Pauli, and tensor network methods \cite{begusic2023fastsim,doi:10.1126/sciadv.adk4321,PhysRevLett.133.120603,Rudolph2023ClassicalSS,PRXQuantum.5.010308,liao2023simulationibmskickedising,PhysRevResearch.6.013326} classically reproduced the $127$-qubit kicked Ising experiment \cite{Kim2023Evidence} that motivated recent claims of quantum utility, in some cases faster than the quantum device and to higher accuracy.

Pauli propagation writes the observable as a sum of Pauli strings and pushes each string through the circuit one gate at a time. Every rotation that does not commute with a string splits it into a cosine and a sine branch, so the number of strings can grow exponentially with depth. Truncation keeps the method practical at a bounded cost in accuracy: it drops the strings whose Pauli weight, the number of non-identity factors, exceeds a maximum weight $w_{\max}$, and those whose coefficients are small (\Cref{sec:truncation}). The truncated operator is the \emph{Pauli propagation surrogate} (\pps) \cite{Rudolph2023ClassicalSS,6vd7-l9bn}. Because its cosine and sine factors are kept as functions of the angles, the surrogate can be stored once and re-evaluated at any angles \cite{Rudolph2023ClassicalSS}.

\paragraph{Existing tools and related methods.}
Pauli propagation has a growing body of guarantees, extensions, and libraries. The low-weight efficient simulation algorithm (LOWESA) gives a provable average-case guarantee for noisy parameterized circuits \cite{Fontana2025Classical} and is reused to build surrogate expectation landscapes \cite{Rudolph2023ClassicalSS}; closely related lines of work appear as Clifford perturbation theory \cite{Begui2025Simulating}, as simulations of sparse Pauli dynamics \cite{begusic2023fastsim,doi:10.1126/sciadv.adk4321}, and as observable backpropagation on Pauli paths, which bounds the truncation error under single-qubit Pauli noise \cite{PhysRevLett.133.120603}. Recent extensions cover arbitrary local noise, imaginary time evolution, quantum enhanced Pauli propagation, and the merging of Pauli strings related by a circuit symmetry \cite{fb28-wlv2,GomezLurbe2026itepp,majumder2026quantumenhancedpaulipropagation,teng2026leveragingsymmetrymergingpauli}, and separate libraries target the processing of Pauli strings at high throughput and time evolution in the Heisenberg picture \cite{kroetz2026paulib,SciPostPhysCodeb.54,SciPostPhysCodeb.54-r1.5}. The guarantees for noisy circuits belong to a broader line of results on the classical tractability of noisy circuits, such as sampling from noisy random circuits \cite{10.1145/3564246.3585234}.

Other Pauli propagation engines exist. The open-source PauliPropagation.jl library propagates the Pauli sum gate by gate on the central processing unit (CPU), merging duplicate strings and truncating after each gate \cite{6vd7-l9bn}; the Qiskit addon pauli-prop \cite{pauli-prop} also runs on the CPU and truncates by a coefficient cutoff and a cap on the number of retained terms; NVIDIA's cuQuantum software development kit (SDK) added cuPauliProp~\cite{cuquantum_cupauliprop}, a graphics processing unit (GPU) Pauli propagation engine, in release 25.11; and BlueQubit's hosted GPU Pauli path simulator has been benchmarked against other simulators \cite{mazumder2026benchmarkingzerosetup}.

\pado\ \emph{compiles} a circuit and its observables: it runs the Heisenberg propagation once, with truncation and zero-filtering, and stores the result as a program of per-gate sparse matrices in which each rotation angle enters only through a cosine and a sine factor supplied at evaluation. The circuit itself is left as written; no mapping to a hardware gate set or qubit layout (transpilation) is involved. Every later query, batched expectation values and gradients, reuses the program, evaluated and differentiated within PyTorch. \pado\ was developed independently of these tools and concurrently with several of them.

The surrogate viewpoint has been used to build classical surrogates of variational landscapes and their quantum enhanced patches \cite{Rudolph2023ClassicalSS,fhc5-8sm6}, to model a trained circuit's outputs as a function of its data inputs \cite{PhysRevLett.131.100803}, and to pre-train circuits through a differentiable symbolic representation \cite{monaco2026symbolicpaulipropagationgradientenabled}. It has also been used to show that families such as quantum convolutional neural networks are effectively classically simulable \cite{8qt9-72ts,Cerezo2025Does}. The surrogate further serves as a classical initializer for variational optimization, even where it is an unreliable simulator \cite{4kyq-n8jb}, and as the classical training stage that finds circuits preparing ground states on one hundred qubits or more, with trained circuits run on hardware at 48 qubits \cite{Lin2026utilityscalequantum}. Other work quantifies \emph{when} Pauli path simulation can be trusted \cite{gharibyan2025practicalguideusingpauli}. Further work makes it usable for optimization with gradients, by sampling Pauli paths for unbiased gradient estimates \cite{zhang2026stochasticpaulipathsimulatorlargescale} or by backpropagating through the Pauli propagation simulation at a memory cost lower than that of conventional reverse-mode automatic differentiation \cite{lin2026backpropagatingpaulipropagation}. \pado\ complements these efforts on the implementation side: it compiles propagation into \emph{sparse linear maps executed in PyTorch}, so that the simulation runs inside a machine learning framework, as tensor network simulation does in \cite{Zhang2023tensorcircuit}.

\begin{figure}[t]
    \centering
    \includegraphics[width=\linewidth]{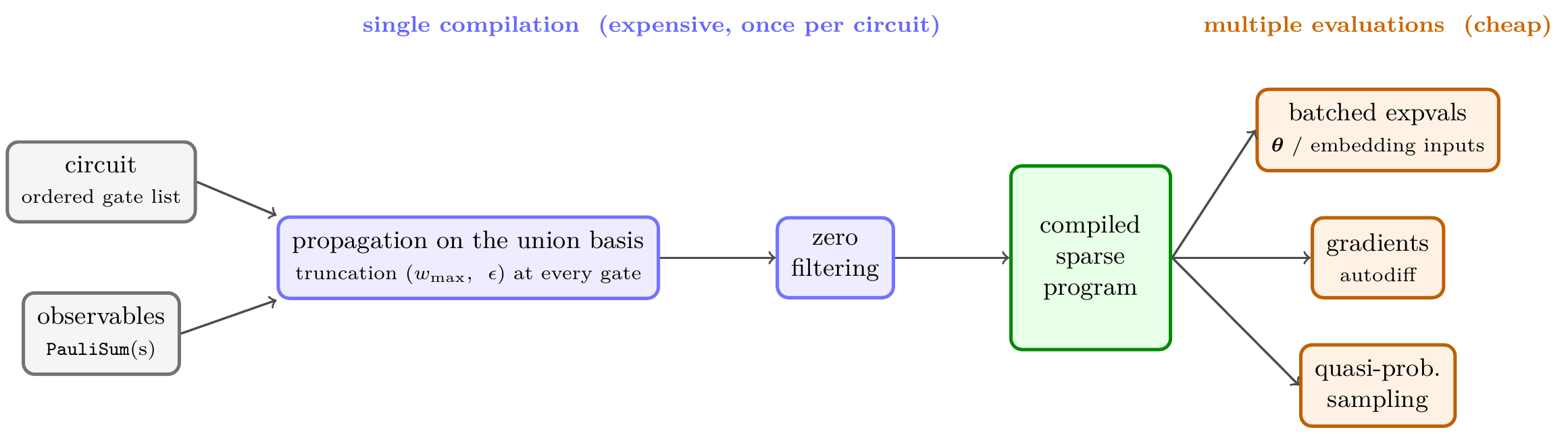}
    \caption{Overview of \pado: a single compilation, with truncation applied at each propagation step, followed by multiple evaluations with gradients.}
    \label{fig:overview}
\end{figure}

\paragraph{Contributions.}
The contributions of this work are the following.
\begin{itemize}
    \item \textbf{Single compilation, multiple evaluations.} A circuit and its observables are propagated, truncated, zero-filtered, and compiled \emph{once} into a single program; every later query (batched expectation values and gradients) reuses it (\Cref{fig:overview}). This matches the pattern of repeated evaluation in variational and quantum machine learning training, which queries the same circuit hundreds of times per run.
    \item \textbf{Zero-filtering of the propagated operator.} After propagation, \pado\ removes every branch that cannot reach a Pauli string with nonzero expectation in the reference state $\ket{0}^{\otimes n}$, so zero-filtering is exact. On the Erd\H{o}s--R\'enyi quantum approximate optimization algorithm (\qaoa) families measured here it shrinks the evaluation working set (the Pauli strings that remain after the last propagation step and zero-filtering) by roughly one to three orders of magnitude (\Cref{sec:zero-filter}).
    \item \textbf{Differentiable execution on GPU tensors.} Pauli strings are stored as symplectic bitmasks (two bits per qubit, \Cref{sec:compile-once}) and propagation is expressed as sparse linear maps with explicit cosine/sine branches, executed as PyTorch tensor operations \cite{10.1145/3620665.3640366}; the rotation angles, trainable or supplied by data, enter those maps as ordinary PyTorch tensors. The compiled program is therefore \emph{differentiable}: \pado\ computes the derivatives of its expectation values with respect to every rotation angle in a backward pass registered with PyTorch's automatic differentiation, so the gradient of a loss built from those values reaches the angles and any PyTorch layer that produced them (\Cref{sec:evaluation}), without parameter-shift re-evaluation.
    \item \textbf{Composable truncation.} Max-weight and coefficient truncation trade accuracy against cost through explicit thresholds; both are checked against exact baselines, max-weight truncation up to $28$ qubits and coefficient truncation at $16$ (\Cref{sec:trunc-cost}).
    \item \textbf{Batched embedding inputs, quasi-probability reconstruction, and reference checks.} \pado\ provides evaluation over the Cartesian product of data and parameter batches, reconstruction of the circuit's output quasi-probabilities from $Z$ correlators (\Cref{sec:api}), and reference helpers that cross-check small systems against exact statevector values.
\end{itemize}
Each of these design choices has a counterpart in some existing engine, but none of the engines compared in \Cref{sec:benchmarks} has all of them: GPU execution, a native backward pass, and batched embedding inputs, on a compiled program that is reused across queries inside PyTorch and whose step matrices are pruned by zero-filtering. That combination targets one regime, repeated evaluation with gradients and batched queries, which is the inner loop of variational and quantum machine learning training. \Cref{sec:benchmarks} records where the compared engines offer their own version of each design choice.

In that comparison \pado\ evaluates up to $10\times$ faster than cuPauliProp on $4$- to $12$-qubit instances. On a $16$-qubit instance in the forward-plus-backward configuration it occupies $3.1\times$ less GPU memory at maximum weight $w_{\max}=6$ (cuPauliProp holding complex coefficients) and completes $w_{\max}=7$, where cuPauliProp exhausts the $80$\,GB NVIDIA A100.\footnote{Throughout the paper, GB means $2^{30}$ bytes and MB means $2^{20}$ bytes.} The compiled program reaches $100$ qubits at $w_{\max}=3$ (\Cref{sec:resource-scaling}).

The rest of the paper is organized as follows. \Cref{sec:background} reviews Pauli propagation, truncation and its error bounds, and noise channels in the Pauli picture. \Cref{sec:details} describes what \pado\ compiles, stores, prunes, and differentiates, with measurements of each design choice within the engine, and documents the interface. \Cref{sec:trunc-cost} measures the accuracy of the compiled program against exact baselines and the cost of building and evaluating it. \Cref{sec:benchmarks} compares \pado\ with other engines on common instances, \Cref{sec:examples} works through four applications, and \Cref{sec:conclusion} states limitations and extensions.

\section{Background}
\label{sec:background}

\subsection{The Pauli propagation surrogate (\pps)}

A \emph{surrogate} is an inexpensive model that stands in for an expensive function. Here the expensive function is the map from the circuit parameters $\bm{\theta}$ to the circuit's expectation values, and the surrogate is a classical object that, once built, returns approximations to those values at any $\bm{\theta}$. \pps\ denotes the surrogate that Pauli propagation yields, the truncated effective observable of the Introduction. Let the target observable be
\begin{equation}
    O = \sum_j c_j P_j,
\end{equation}
where each $P_j$ is a Pauli string, a tensor product of single-qubit Pauli matrices. Pauli propagation pushes $O$ through the circuit gate by gate, from the last gate to the first:
\begin{equation}
    O_{\mathrm{eff}}(\bm{\theta}) = U^\dagger(\bm{\theta}) O U(\bm{\theta}).
\end{equation}
The effective observable is represented as a sparse sum of Pauli strings together with the cosine/sine factors needed to regenerate the coefficients at any $\bm{\theta}$.

\pado\ implements the \pps\ workflow with the structure of \Cref{fig:overview}, a single compilation followed by multiple evaluations:
\begin{enumerate}
    \item express the cost or Hamiltonian as a Pauli sum,
    \item propagate each term through the circuit in the Heisenberg picture,
    \item truncate intermediate growth using rules based on structure and on coefficients,
    \item evaluate the compiled program for many parameter values.
\end{enumerate}

Steps 2--3 are what \pado\ compiles into sparse tensor operations, and step 4 runs the result (\Cref{sec:details}).

\paragraph{Pauli propagation mechanics.}
The method's cost is set by the propagation, step~2 of the workflow above. With $O$ a weighted sum of Pauli strings, computing $U^\dagger O U$ reduces to repeatedly rewriting Pauli strings under gate conjugation.

In this picture Clifford gates induce a one-to-one mapping on Pauli strings and therefore do not themselves create branching. The main source of complexity is a parameterized non-Clifford operation such as a Pauli rotation
\begin{equation}
    R_G(\theta) = e^{-i \theta G / 2}.
\end{equation}
Its action on a Pauli string $P$ depends on the commutation relation between $P$ and the generator $G$:
\begin{equation}
    R_G^\dagger(\theta)\, P \, R_G(\theta)
    =
    \begin{cases}
        P, & [P,G]=0,\\
        \cos(\theta) P + \sin(\theta) P', & \text{otherwise},
    \end{cases}
    \label{eq:rot-rule}
\end{equation}
where $P' = i[G,P]/2$ is another Pauli string generated by the non-commuting branch.

For example, let the circuit apply a CNOT from qubit $0$ to qubit $1$ and then $R_X(\theta)$ on qubit $0$, and let $O=Z_0$. Propagation starts at the last gate: $Z_0$ does not commute with $X_0$, so the rotation gives $\cos\theta\,Z_0+\sin\theta\,Y_0$; the CNOT then maps each string to one string, $Z_0\mapsto Z_0$ and $Y_0\mapsto Y_0X_1$, giving $O_{\mathrm{eff}}=\cos\theta\,Z_0+\sin\theta\,Y_0X_1$. In the state $\ket{0}^{\otimes 2}$ only the first string has nonzero expectation, so $\langle Z_0\rangle=\cos\theta$.

\begin{figure}[t]
    \centering
    \includegraphics[width=0.9\linewidth]{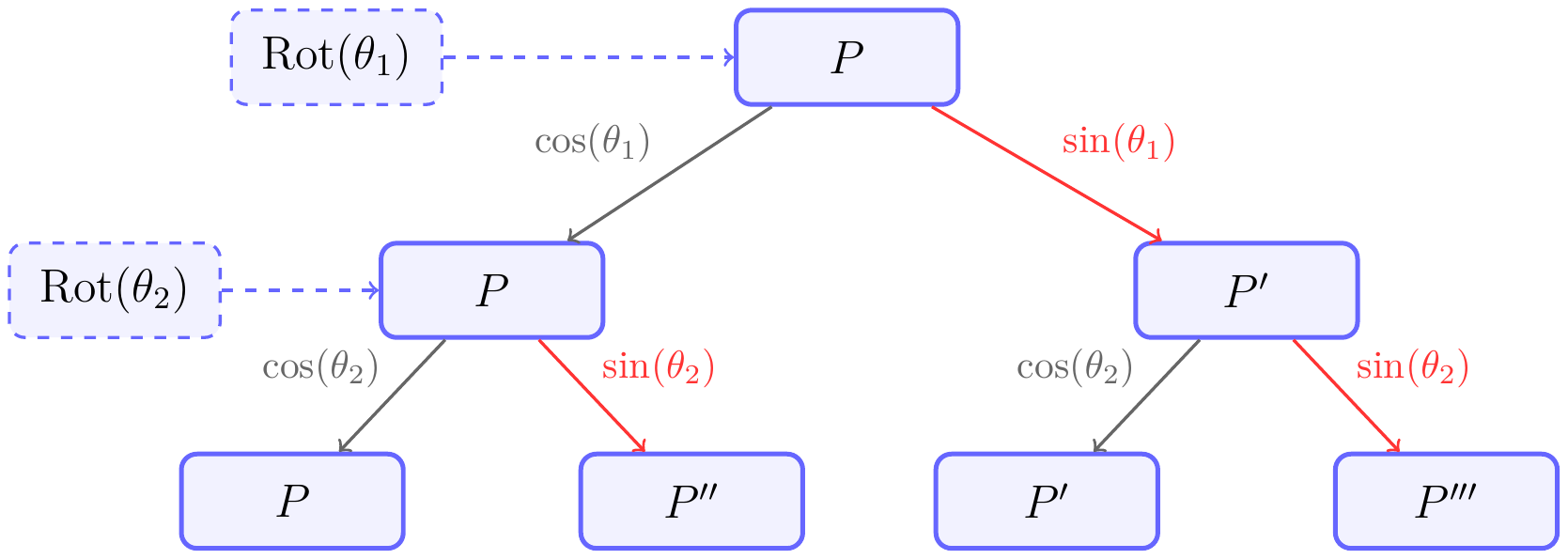}
    \caption{Conceptual branching tree for Pauli propagation under parameterized rotations: non-commuting gates split a Pauli string into cosine and sine branches. $\mathrm{Rot}(\theta_\ell)$ is the rotation $R_G(\theta_\ell)$ of \Cref{eq:rot-rule}, and $P'$, $P''$, $P'''$ the distinct strings its successive sine branches produce.}
    \label{fig:pauli_tree}
\end{figure}

A single initial string can generate up to $2^m$ branches, where $m$ is the number of non-commuting rotations met during propagation (\Cref{fig:pauli_tree}). Different paths may produce the same Pauli string, whose coefficients are then summed.

\subsection{Truncation strategies and error bounds}
\label{sec:truncation}

\paragraph{Max-weight truncation.}
Max-weight truncation discards terms whose Pauli weight (number of non-identity factors) exceeds $w_{\max}$ \cite{Rudolph2023ClassicalSS,6vd7-l9bn}. The motivation is the expectation value taken at the end of propagation: for $\rho_0=\ket{0}\!\bra{0}^{\otimes n}$, a string with $W$ non-identity factors drawn uniformly from $\{X,Y,Z\}$ has $\Tr[\rho_0 P]=0$ with probability $1-(1/3)^W$, since $\Tr[\rho_0 P]\neq0$ requires all $W$ factors to be $Z$, so under this model strings of high weight rarely contribute \cite{6vd7-l9bn}. The rule depends only on the Pauli support and is therefore independent of the rotation angles.

\paragraph{Weighted (per-axis) max-weight truncation.}
\pado\ also offers a weighted variant: each Pauli axis is assigned a per-axis weight $w_X$, $w_Y$, $w_Z$, all $1$ by default and set through the presets of \Cref{sec:compile-once}, and a string is kept only if the per-axis weights of its non-identity factors sum to at most $w_{\max}$, $\sum_{q} w_{P_q}\le w_{\max}$.

\paragraph{Coefficient truncation.}
\emph{Coefficient truncation} discards terms whose absolute coefficient falls below a threshold $\epsilon$, $|c_j|<\epsilon$ \cite{6vd7-l9bn}. Unlike max-weight truncation it depends on the parameter values.

\paragraph{Other structural filters.}
Other filters in the literature cap a quantity accumulated along each path, such as the trigonometric \emph{frequency} (the number of cosine or sine factors a path picks up \cite{Fontana2025Classical,6vd7-l9bn}).

\paragraph{Truncation error bounds.}
Because $|\Tr[P_j\rho]|\le1$ for every Pauli string and state, a single truncation that drops a set $T$ of terms from $O=\sum_j c_j P_j$, leaving $\hat O$, changes an expectation value by at most the discarded coefficient mass,
\begin{equation}
    \bigl|\Tr[(O-\hat O)\rho]\bigr| \;\le\; \sum_{j\in T} |c_j| \;=:\; \Delta,
    \label{eq:trunc-l1}
\end{equation}
and the total error is at most the sum of the discarded masses of the individual truncations,
$\Delta_{\mathrm{tot}}\le\sum_{\ell}\Delta_\ell$, with no assumption on the state or the circuit \cite{6vd7-l9bn}.

Truncation is thus a controllable approximation; \Cref{sec:trunc-accuracy} measures its effect.

\subsection{Noise channels in the Pauli picture}
\label{sec:noise}

\paragraph{Pauli noise.}
Pauli noise (homogeneous or inhomogeneous dephasing and depolarizing) is the family of unital channels diagonal in the Pauli basis. For the single-qubit inhomogeneous depolarizing channel with $X/Y/Z$ error probabilities $p_x,p_y,p_z$,
\begin{equation}
    \mathcal{E}(\rho)=p_I\,\rho+p_x X\rho X+p_y Y\rho Y+p_z Z\rho Z,\qquad p_I=1-(p_x+p_y+p_z),
\end{equation}
the corresponding Pauli transfer action (identical in the Schr\"odinger and Heisenberg pictures) leaves the identity fixed and damps each single-qubit Pauli,
\begin{equation}
    \mathcal{E}[I]=I,\quad
    \mathcal{E}[X]=(1-2p_y-2p_z)\,X,\quad
    \mathcal{E}[Y]=(1-2p_x-2p_z)\,Y,\quad
    \mathcal{E}[Z]=(1-2p_x-2p_y)\,Z.
    \label{eq:depol-ptm}
\end{equation}
The Pauli set is therefore unchanged: these channels do not branch and only multiply each coefficient by a factor of magnitude at most one, since each factor is $1-2(p_a+p_b)$ with $0\le p_a+p_b\le p_x+p_y+p_z\le1$.

\paragraph{Amplitude damping.}
In the Heisenberg picture one propagates the dual map $\mathcal{N}^\dagger$ of a channel $\mathcal{N}$. The dual map is always unital, so even non-unital noise never branches on the identity. For amplitude damping with damping rate $\gamma$ the only branching is $\mathcal{N}^\dagger[Z]=\gamma I+(1-\gamma)Z$, while $X$ and $Y$ are only damped \cite{j1gg-s6zb}.

\paragraph{Noise and truncation error.}
Each Pauli noise layer multiplies the coefficient of every string with support on the noisy site, and with it any discarded mass $\Delta$ of \Cref{eq:trunc-l1}, by a factor of magnitude at most one; strings of high weight, damped at every non-identity site when $p_x,p_y,p_z>0$, are suppressed fastest. Noise thus reinforces the structure of low weight and small coefficients that truncation exploits; this is the mechanism behind the average-case guarantees for circuits with Pauli noise \cite{Fontana2025Classical,PhysRevLett.133.120603,10.1145/3564246.3585234}, and any incoherent local noise, including non-unital noise, truncates a typical circuit to an effective logarithmic depth \cite{fb28-wlv2,Mele2026Noise}. At a fixed term budget noise should therefore lower the truncation error; \Cref{sec:frontier} checks this.

\section{The \pado\ Engine: Design and Interface}
\label{sec:details}

This section describes how \pado\ realizes the design choices of the Introduction and measures each one within the engine, by switching it off or by changing the preset, so that its effect is separated from the engine-to-engine differences of \Cref{sec:benchmarks}. \Cref{sec:compile-once} defines the compiled program and the stages that build it, \Cref{sec:zero-filter} the zero-filtering pass, \Cref{sec:evaluation} evaluation, batching, and differentiation in PyTorch, and \Cref{sec:api} the interface and the supported gate set.

\subsection{Single compilation, multiple evaluations}
\label{sec:compile-once}

\paragraph{Data structure.}
The compiled program is a list of per-gate sparse matrices together with the Pauli strings of its output rows. Its logical content is
\begin{equation}
    \mathcal{P}
    =
    \bigl(
    n,\ X,\ Z,\ \mathbf{c}_{\mathrm{init}},\ \{\mathcal{S}_\ell\}_{\ell=1}^{L}
    \bigr),
\end{equation}
where $n$ is the qubit count. Here, $X$ and $Z$ are mask tensors listing the Pauli strings present after propagation through every gate, the propagated basis (after zero-filtering, only its $\{I,Z\}^{\otimes n}$ rows). $\mathbf{c}_{\mathrm{init}}$ is a coefficient vector on the union basis, the union of the observables' Pauli strings; the compiler uses it only to estimate coefficient magnitudes for coefficient truncation. Each $\mathcal{S}_\ell$ is one compiled propagation step, the sparse matrices of one gate. The intermediate bases are not stored; they appear only as the row and column dimensions of the steps.

The mask representation is the binary symplectic encoding of Pauli strings, with one bit pair per qubit as in \cite{PhysRevA.70.052328}, as used, for example, in \cite{kroetz2026paulib}. For row $i$ and qubit $q$, the bit pair $(x_{iq}, z_{iq})$ determines the local operator:
\[
    (0,0)\mapsto I,\qquad
    (1,0)\mapsto X,\qquad
    (0,1)\mapsto Z,\qquad
    (1,1)\mapsto Y.
\]
Each active Pauli string occupies one row of the paired mask tensors. Two bits per qubit put a string at $16$ bytes for $n\le63$, which keeps the propagated basis, millions of strings before zero-filtering (\Cref{tab:zf-ablation}), within memory. Whether a string $(x,z)$ commutes with a gate generator $(x_G,z_G)$, the test that decides branching in \Cref{eq:rot-rule}, is the parity of the set bits of $(x\wedge z_G)\oplus(z\wedge x_G)$, so the C\texttt{++} extension answers it with a few word operations per string. A mask uses $63$ bits of each $64$-bit integer word and spans several words when $n>63$, so the qubit count is not limited by the word size; the $100$-qubit programs of \Cref{sec:resource-scaling} use two words per mask.

\paragraph{Sparse tensor formulation and GPU execution.}
\pado\ formulates propagation as linear algebra on the coefficients of the active Pauli strings. If $N_\ell$ denotes the number of unique active strings before step $\ell$, which truncation keeps far below the $4^n$ Pauli strings on $n$ qubits, then the propagated observable at that stage is represented by a dense coefficient vector $\mathbf{v}_\ell \in \mathbb{R}^{N_\ell}$. Applying the next gate induces a sparse transition map with at most two nonzero entries per column, since a gate maps a string to at most two strings (\Cref{eq:rot-rule}); this is the matrix applied by the compiled step $\mathcal{S}_\ell$ above,
\begin{equation}
    \mathbf{M}_\ell(\theta_\ell) \in \mathbb{R}^{N_{\ell+1} \times N_\ell},
\end{equation}
whose nonzero pattern is determined by the Pauli rewrite rules and whose values depend on the gate parameter when the gate is non-Clifford. The full Heisenberg evolution through $L$ steps can then be written as
\begin{equation}
    \mathbf{v}_{\mathrm{out}}
    =
    \mathbf{M}_{L}(\theta_L)\,\mathbf{M}_{L-1}(\theta_{L-1})\cdots\mathbf{M}_{1}(\theta_1)\,\mathbf{v}_{\mathrm{in}}.
\end{equation}
Here $\mathbf{v}_{\mathrm{in}}$ is the coefficient vector of an observable on the union basis (a column of $V_0$, \Cref{sec:evaluation}). The steps are numbered in propagation order, one per gate, so $\mathbf{M}_1$ applies the last gate of the circuit and $\mathbf{M}_L$ the first; evaluation runs this product in transposed form (\Cref{sec:evaluation}). Propagation is then a sequence of products of a sparse matrix with a vector or a matrix, operations for which GPU kernels have been developed \cite{9355309}.

\paragraph{Separation of parameters from structure.}
\Cref{eq:rot-rule} sorts the strings entering a rotation step into two groups: a string that commutes with the generator passes through unchanged, and every other string contributes $\cos\theta_\ell$ times itself and $\sin\theta_\ell$ times the string $P'$ of \Cref{eq:rot-rule}. A compiled step therefore represents its action as three sparse components,
\begin{equation}
    \mathbf{v}_{\ell+1}
    =
    M_{\mathrm{const}}^{(\ell)} \mathbf{v}_\ell
    +
    \cos(\theta_\ell)\, M_{\cos}^{(\ell)} \mathbf{v}_\ell
    +
    \sin(\theta_\ell)\, M_{\sin}^{(\ell)} \mathbf{v}_\ell,
    \label{eq:step-decomp}
\end{equation}
where $M_{\mathrm{const}}$ carries the commuting strings, $M_{\cos}$ the non-commuting strings copied in place, and $M_{\sin}$ their sine branches; a Clifford or noise step has the first component only. The angle enters \Cref{eq:step-decomp} only through the two scalars $\cos\theta_\ell$ and $\sin\theta_\ell$, so the three matrices, the \emph{step matrices}, are fixed at compilation. Three consequences follow, and \Cref{sec:evaluation} builds on each. The same matrices serve every parameter vector, so a program is compiled once and queried many times. A batch of parameter vectors or embedding inputs changes only the scalars, so it is evaluated with the scalar factors carried along a batch dimension while each matrix is read once. The derivative with respect to $\theta_\ell$ is $-\sin(\theta_\ell)\,M_{\cos}^{(\ell)}\mathbf{v}_\ell+\cos(\theta_\ell)\,M_{\sin}^{(\ell)}\mathbf{v}_\ell$, the same two products with the scalars exchanged and one sign flipped, so a gradient is another pass over the same matrices and does not propagate the circuit again.

For a rotation step, $M_{\mathrm{const}}$ and $M_{\cos}$ are stored together as one index map. For each output row whose string the gate leaves in place, the map gives the input row it copies and whether the copy is multiplied by $\cos\theta_\ell$. Sine branches are stored as explicit sparse entries, on a new row for a new string or on the existing row of a string already present. The step's linear operator is the same regardless of how its entries are stored. Clifford and noise steps store $M_{\mathrm{const}}$ alone.

\paragraph{Propagation pipeline.}
Given a circuit $U = U_L \cdots U_2 U_1$ and an observable $O$, the compiled program is built by applying the gate updates in reverse order,
\begin{equation}
    O^{(k-1)} = U_k^\dagger O^{(k)} U_k,
    \qquad
    O^{(L)} = O,
\end{equation}
so that step $\ell$ of the propagation order conjugates by the gate $U_{L+1-\ell}$. The compile path runs this recursion in three stages, which take the circuit, the observables, and the preset and return the program $\mathcal{P}$ above.
\begin{enumerate}
    \item \textbf{Union basis.} The compiler collects the Pauli strings of all observables into one ordered list, the union basis, and records on it the coefficient vector $\mathbf{c}_{\mathrm{init}}$ that coefficient truncation reads. All observables are then propagated together through one step sequence, and their own coefficients enter only at evaluation, as the columns of $V_0$ (\Cref{sec:evaluation}), so a Hamiltonian with many terms or a list of observables costs one propagation.
    \item \textbf{Propagation.} One pass over the gates in reverse order turns the union basis into the step sequence $\{\mathcal{S}_\ell\}$. At each gate the C\texttt{++} extension (\Cref{sec:api}) rewrites the masks of the active strings, generates the sine branches of a rotation, merges the rows that now hold the same string, and then applies the truncations to the merged rows, max-weight ($w_{\max}$) and, when a threshold $\epsilon$ is given, coefficient truncation; a row that fails a rule is dropped inside the extension and never reaches Python. The pass emits the three components of \Cref{eq:step-decomp} for the step; the strings active before it become the column dimension of this step and the row dimension of the next, and the intermediate bases themselves are not stored.
    \item \textbf{Zero-filtering.} The compiler propagates a survival mask (\Cref{sec:zero-filter}) from step $L$ to step $1$, starting at the output rows in $\{I,Z\}^{\otimes n}$, and prunes the sparse steps, the propagated basis, the union basis, and $\mathbf{c}_{\mathrm{init}}$ to the rows it retains. Propagation generates every reachable string; this stage removes the strings that cannot contribute to an expectation value in $\ket{0}^{\otimes n}$, so evaluation runs on the retained rows only.
\end{enumerate}
The propagation stage takes most of the compile time, and zero-filtering costs no more than the propagation it follows (\Cref{sec:resource-scaling}).

\paragraph{Circuit and observables.}
The compiler takes a circuit as an ordered list of gate objects applied left to right: Pauli rotations $e^{-i\theta P/2}$ with an arbitrary Pauli string $P$ as generator, the Clifford gates of \Cref{sec:scope}, and the two noise channels of \Cref{sec:noise}. Each rotation names its angle by an index: \texttt{param\_idx} marks a trainable angle, and \texttt{embedding\_idx} marks a data input, an embedding, supplied per sample at evaluation and not trained (\Cref{sec:evaluation}). An observable is a \texttt{PauliSum}, Pauli strings with coefficients; \Cref{lst:quickstart} shows the tuple forms that build one.

A circuit is usually written with the \texttt{Circuit} builder, which wraps these objects in methods that can be chained and accumulates the gate list as \texttt{qc.gates}; \texttt{qc.compile(observables=..., preset=...)} and \texttt{qc.expvals(thetas)} forward to the functional entry points below.

\paragraph{Compilation.}
\path{compile_program} runs the three stages above and returns a \texttt{CompiledProgram}. Max-weight truncation is always part of the propagation stage; its threshold $w_{\max}$ comes from the preset (the \texttt{max\_weight} field), and every preset of \Cref{tab:presets} sets it above any reachable weight, so a program is untruncated unless \texttt{max\_weight} is given. Storage and compute devices and precision are selected through the same named \emph{preset}, described below, optionally refined with overrides.

The compiler also accepts coefficient truncation: terms whose absolute coefficient at one fixed parameter vector, the \emph{build point} \path{build_thetas}, falls below the threshold $\epsilon$, set by \path{build_min_abs}, are dropped. With several terms or observables the threshold is applied to the observable on the union basis, whose coefficient vector $\mathbf{c}_{\mathrm{init}}$ holds, for each string, the largest $\lvert c\rvert$ that string has in any listed observable (\Cref{app:internals}). A rotation driven by an embedding input reads its build angle from \path{build_thetas} at its \texttt{embedding\_idx}, so \path{build_thetas} must cover the embedding indices too.

\paragraph{Presets.}
A preset (\texttt{ExecutionPreset}) is an immutable record of where the compiled program is kept, where it is evaluated, the coefficient \texttt{dtype}, the truncation fields $w_{\max}$ and its per-axis weights, and the block size \texttt{chunk\_size}. The three presets are listed in \Cref{tab:presets}: \texttt{cpu} keeps program and evaluation on the CPU, which has the larger memory; \texttt{gpu} keeps both on the faster GPU; \texttt{hybrid} keeps the program on the CPU and evaluates on the GPU, copying a step matrix of at most \texttt{chunk\_size} rows to the GPU once and caching it, and a larger one in blocks of \texttt{chunk\_size} rows on every call ($5\times10^{7}$ for \texttt{hybrid}, $10^{10}$ for \texttt{cpu} and \texttt{gpu}, which never split). All default to \texttt{float64}, with \texttt{float32} selected explicitly for runs bound by memory or throughput. Passing \texttt{preset\_overrides} to \path{compile_program} customizes a preset; \path{resolve_preset(name, overrides=...)} returns the resulting record for inspection.

\begin{table}[t]
    \centering
    \small
    \begin{tabular}{l l l}
        \toprule
        Preset & Memory device & Compute device \\
        \midrule
        \texttt{cpu}    & \texttt{cpu} & \texttt{cpu} \\
        \texttt{gpu}    & \texttt{cuda} & \texttt{cuda} \\
        \texttt{hybrid} & \texttt{cpu} & \texttt{cuda} \\
        \bottomrule
    \end{tabular}
    \caption{Execution presets: where each stores and where it evaluates the compiled program. \texttt{cuda} is the PyTorch device string for the GPU of the installed build.}
    \label{tab:presets}
\end{table}

\paragraph{Evaluation.}
\texttt{program.expvals(thetas)} returns one expectation value per observable; repeated calls reuse the compiled program and re-run only the evaluation of \Cref{sec:evaluation}.

The minimal workflow of \Cref{lst:quickstart} combines these steps. It builds a circuit on six qubits with an $R_X$ rotation on every qubit, an $R_{ZZ}$ rotation on every neighboring pair, and a final CNOT, each rotation with its own trainable angle. It declares three observables: $Z_0$, the correlator $Z_0Z_5$, and the Ising chain energy $\sum_{q=0}^{4}Z_qZ_{q+1}$, the last passed as a list of terms so that it counts as one observable. It compiles them once under the \texttt{gpu} preset and evaluates them at one parameter vector, which returns three values; a call with another parameter vector reuses the same program. The last two lines build the same program from the gate list without the \texttt{Circuit} builder.

\begin{lstlisting}[caption={Minimal workflow, a single compilation followed by multiple evaluations (condensed).},label={lst:quickstart}]
import torch
from padopauli import Circuit, compile_program

n = 6
# 1) Build a parameterized circuit; every gate call returns the circuit.
qc = Circuit(n)
pidx = 0
for q in range(n):
    qc.rx(q, param_idx=pidx); pidx += 1
for q in range(n - 1):
    qc.rzz(q, q + 1, param_idx=pidx); pidx += 1
qc.cnot(0, 1)

# 2) Observables are (pauli, qubits[, coeff]) tuples; a LIST of tuples is one
#    observable with several terms, not several observables.
observables = [
    ("Z", [0]),                                        # <Z_0>
    ("ZZ", [0, n - 1]),                                # end-to-end correlator
    [("ZZ", [q, q + 1], 1.0) for q in range(n - 1)],   # an Ising-chain energy
]

# 3) Compile once, then evaluate for many parameter vectors.
qc.compile(observables=observables, preset="gpu", dtype="float64")
thetas = torch.linspace(0.1, 1.2, qc.n_params, dtype=torch.float64)
vals = qc.expvals(thetas)   # shape: (3,), one value per observable

# Without the builder: the gate list directly, plus the register width.
program = compile_program(circuit=qc.gates, observables=observables,
                          n_qubits=n, preset="gpu")
\end{lstlisting}

\subsection{Zero-filtering}
\label{sec:zero-filter}

After propagation, zero-filtering removes from the compiled program every term that cannot contribute to an expectation value in the reference state. Unlike the truncations of \Cref{sec:truncation}, it introduces no approximation: every removed term contributes zero at every parameter value. The reference state is $\rho_0 = \ket{0}\!\bra{0}^{\otimes n}$. For such a state, any Pauli string containing at least one $X$ or $Y$ factor has zero expectation value, so only strings in $\{I,Z\}^{\otimes n}$ contribute:
\begin{equation}
    \Tr[\rho_0 P_i]
    =
    \begin{cases}
        1, & P_i \in \{I,Z\}^{\otimes n},\\
        0, & \text{otherwise}.
    \end{cases}
    \label{eq:zero-overlap}
\end{equation}

\begin{figure}[t]
    \centering
    \includegraphics[width=0.62\linewidth]{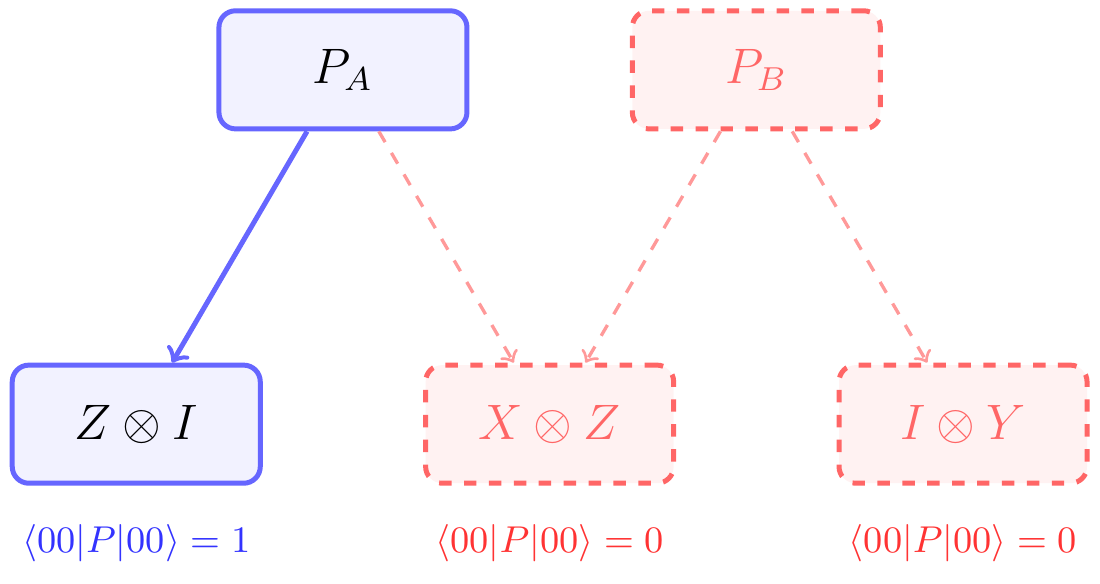}
    \caption{Zero-filtering at compile time. Solid blue, strings reaching a kept output row; dashed red, strings removed because $\Tr[\rho_0 P]=\langle00|P|00\rangle=0$ for them or for every string they produce.}
    \label{fig:zero_filtering}
\end{figure}

If \Cref{eq:zero-overlap} is applied only to the output rows, every evaluation still propagates the branches that lead only to strings with $\Tr[\rho_0 P]=0$, and their contribution is then multiplied by zero. Zero-filtering removes these branches at compile time with one rule, applied from step $L$ to step $1$: a string before step $\ell$ is kept if and only if at least one of the strings it produces in step $\ell$ is kept. \Cref{fig:zero_filtering} applies the rule to one step. $P_A$ is kept because it produces $Z\otimes I$, for which $\Tr[\rho_0 P]=1$. $P_B$ is removed because both strings it produces have $\Tr[\rho_0 P]=0$. The branch from $P_A$ to $X\otimes Z$ is removed as well, since $X\otimes Z$ is not kept.

In the notation of \Cref{sec:compile-once}, the rule is a recursion on the \emph{survival mask} $\mathbf{m}_\ell\in\{0,1\}^{N_\ell}$, which marks the strings kept before step $\ell$. On the output rows the mask is \Cref{eq:zero-overlap} itself, $[\mathbf{m}_{L+1}]_r=\Tr[\rho_0 P_r]$, and the earlier masks follow as
\begin{equation}
    \mathbf{m}_{\ell}
    =
    \operatorname{sgn}\!\Bigl(\bigl(|M_{\mathrm{const}}^{(\ell)}|+|M_{\cos}^{(\ell)}|+|M_{\sin}^{(\ell)}|\bigr)^{\top}\mathbf{m}_{\ell+1}\Bigr),
    \qquad \ell = L, L-1, \dots, 1,
    \label{eq:zf-mask}
\end{equation}
where ${}^{\top}$ denotes the transpose, which carries a mask on the strings after step $\ell$ back to the strings before it, and $|\cdot|$ is the entrywise absolute value. The nonzero entries of the matrix in parentheses mark the pairs of strings that step $\ell$ connects; the engine forms it as a boolean OR over the nonzero patterns of the three components, with the index map standing in for $M_{\mathrm{const}}$ and $M_{\cos}$ of a rotation step.

The mask depends only on which strings each step connects, and this makes the pruning exact. The absolute values remove the signs of Pauli conjugation, and taking the components separately removes the factors $\cos\theta_\ell$ and $\sin\theta_\ell$ of \Cref{eq:step-decomp}, so the matrix in parentheses does not depend on $\bm{\theta}$. With either left in, the product in \Cref{eq:zf-mask} could vanish for particular values of $\bm{\theta}$, for example where $\cos\theta_\ell=0$, and remove a string that contributes at other values. A string with $[\mathbf{m}_{\ell}]_k=0$ has no path of any sign to an output row with $\Tr[\rho_0 P_r]=1$, so its contribution is zero for every $\bm{\theta}$, and it is removed before evaluation.

Earlier work notes that, for $\rho_0$, the graph representation of the expectation value ``can be re-evaluated only on the paths leading to'' the strings in $\{I,Z\}^{\otimes n}$ \cite{8qt9-72ts}, and PauliPropagation.jl provides this reduction for its surrogate as \path{zerofilter!} \cite{paulipropagation_jl_repo}. In \pado\ the program is a sequence of sparse matrices that every evaluation reads in full, so the reduction is applied to the matrices themselves: the recursion of \Cref{eq:zf-mask} removes the strings and entries that do not lead to a kept output row at compile time, as a default stage of \path{compile_program}.

On the Erd\H{o}s--R\'enyi \qaoa\ families measured here the pass leaves an evaluation working set roughly one to three orders of magnitude smaller than the propagated basis, and the reduction widens with $n$ on the family of \Cref{tab:zf-ablation}, whose qubit sweep is in \Cref{sec:resource-scaling}.

\subsection{Evaluation, batching, and differentiation}
\label{sec:evaluation}

\paragraph{Evaluation and differentiation.}
At a parameter vector $\bm{\theta}$, \pado\ starts from the survival mask $\mathbf{m}_{L+1}$ of \Cref{sec:zero-filter}, which marks the output rows with $\Tr[\rho_0 P_r]=1$, and multiplies it by the transposed step matrices from step $L$ to step $1$. The result is the adjoint vector of adjoint differentiation for circuit simulators \cite{jones2020efficientcalculationgradientsclassical},
\begin{equation}
    \mathbf{w}(\bm{\theta})
    =
    \mathbf{M}(\bm{\theta})^{\top}\mathbf{m}_{L+1},
\end{equation}
where $\mathbf{M}(\bm{\theta})$ denotes the full compiled sparse map. Since the step matrices do not depend on the parameters (the angles enter only as scalar $\cos/\sin$ multipliers after the sparse multiply), \pado\ transposes and caches them once and reuses them on every call. The \texttt{hybrid} preset is the exception: step matrices with more rows than its block size are streamed in blocks and transposed on every call (\Cref{sec:compile-once}).

For $N_{\mathrm{obs}}$ observables sharing the same compiled union basis, \pado\ assembles their coefficients into a dense matrix
\begin{equation}
    V_0 \in \mathbb{R}^{N_{\mathrm{in}}\times N_{\mathrm{obs}}},
\end{equation}
where $N_{\mathrm{in}}$ is the number of rows of the union basis that survive zero-filtering,
and then recovers all $N_{\mathrm{obs}}$ expectation values by one final dense multiplication
\begin{equation}
    \mathbf{E}(\bm{\theta}) = \mathbf{w}(\bm{\theta})^{\top} V_0 \in \mathbb{R}^{1\times N_{\mathrm{obs}}}.
\end{equation}
The $N_{\mathrm{obs}}$ observables thus share one sparse propagation through the circuit, and their coefficients enter only through the columns of $V_0$.

\paragraph{PyTorch integration and automatic differentiation.}
The evaluation runs in PyTorch and stays inside its computational graph.

If $\mathcal{L}$ denotes a scalar training objective derived from the program's output $\mathbf{v}_{\mathrm{out}}$, then gradients with respect to circuit parameters take the form
\begin{equation}
    \frac{\partial \mathcal{L}}{\partial \theta_k}
    =
    \frac{\partial \mathcal{L}}{\partial \mathbf{v}_{\mathrm{out}}}
    \frac{\partial}{\partial \theta_k}
    \left[
    \mathbf{M}_{L}(\theta_L)\cdots\mathbf{M}_{1}(\theta_1)\,
    \mathbf{v}_{\mathrm{in}}
    \right].
\end{equation}
\paragraph{Two differentiation modes.}
\pado\ offers two implementations of this gradient, selected by a \texttt{diff\_mode} switch. The default, \texttt{vjp}, wraps this evaluation in a custom autograd primitive. The forward pass runs detached and saves, besides its inputs, only the adjoint vectors entering the \emph{parameterized} steps and the $\cos/\sin$ factors it already formed. A backward pass then replays the steps against them, accumulating $\partial\mathcal{L}/\partial\theta_k$ through $-\sin(\theta_k)(\cdot)+\cos(\theta_k)(\cdot)$. The first forward-plus-backward pass also converts the step matrices to the sparse form the backward replay reads and caches them, separately from the forward cache above; the \emph{first} times of \Cref{tab:zf-ablation} include this build. The alternative, \texttt{autograd}, lets PyTorch trace the chain of sparse matrix products and retain its full autograd graph. A backward pass for Pauli propagation in place of a traced graph was reported concurrently and independently in \cite{lin2026backpropagatingpaulipropagation}.

\paragraph{Gradient correctness.}
The reference for \pado's gradient is a central finite difference of the compiled program's own objective. A parameter-shift rule on the circuit would estimate the gradient of the exact circuit instead. On a $12$-qubit, depth-$2$ MaxCut-\qaoa\ instance (Erd\H{o}s--R\'enyi graph, edge probability $0.4$, seed $7$; $78$ parameters drawn uniformly from $[-\pi,\pi]$ with \texttt{numpy.random.default\_rng(123)}; $w_{\max}=5$, \texttt{float64}, step $\delta=10^{-5}$) the analytic gradient agrees with the finite difference to $4.3\times10^{-11}$ over all $78$ components, at the level of the finite-difference error in double precision, and \path{vjp} and \path{autograd} agree to $\sim\!10^{-16}$. Any remaining discrepancy from the exact circuit's gradient is truncation error.

\paragraph{Batched embedding inputs and parameter batches.}
\pado\ assembles the embedding inputs of a dataset and the parameters into batches of angles and evaluates them together on the compiled program.

An embedding batch of size $B_e$ and a parameter batch of size $B_p$ supplied together are evaluated as a Cartesian product in one pass, giving an output of shape $(B_e, B_p, N_{\mathrm{obs}})$. Each embedding value supplies the $\cos/\sin$ factors of its rotation step, so the output is a \emph{nonlinear} function of the embedding angles~\cite{PhysRevA.103.032430}; the $B_e$ inputs enter through the batch dimension of the sparse matrix products before the dense $\mathbf{w}^{\top}V_0$ readout. \Cref{sec:resource-scaling} quantifies the gain over calls with a single vector within the engine, and \Cref{sec:benchmarks} the gain against the per-sample loops of other engines.

In \Cref{lst:hybrid} a \texttt{torch.nn.Linear} maps two features to the four embedding angles of a two-qubit program and one optimizer trains both groups. The gradient on the eight classical weights agrees with a central difference of the loss to $1.2\times10^{-10}$, so no weight is disconnected, and $300$ steps take the binary cross-entropy from $1.709$ to $0.245$ on a task like XOR ($256$ points).

\begin{lstlisting}[caption={A classical layer upstream of the compiled program, trained jointly.},label={lst:hybrid}]
qc = Circuit(n_qubits=2)
qc.ry(0, embedding_idx=0); qc.ry(1, embedding_idx=1)
qc.cnot(0, 1)
qc.ry(0, embedding_idx=2); qc.ry(1, embedding_idx=3)   # the same features enter again
qc.ry(0, param_idx=0);     qc.ry(1, param_idx=1)
qc.compile(observables=[("ZZ", [0, 1])], preset="gpu")

lin    = torch.nn.Linear(2, 4).to(device="cuda", dtype=torch.float64)
thetas = torch.nn.Parameter(torch.zeros(2, dtype=torch.float64, device="cuda"))
opt    = torch.optim.Adam([{"params": lin.parameters()}, {"params": [thetas]}], lr=0.05)

for step in range(300):
    opt.zero_grad(set_to_none=True)
    expv = qc.expvals(thetas, embedding=lin(X))[:, 0]   # classical layer feeds the angles
    prob = torch.clamp((expv + 1) * 0.5, 1e-6, 1 - 1e-6)
    loss = torch.nn.functional.binary_cross_entropy(prob, y)
    loss.backward(); opt.step()                         # reaches lin.weight and thetas
\end{lstlisting}

\paragraph{Example: data embedding and batched inputs.}
\label{sec:ex-embed}
Quantum machine learning models separate \emph{data} inputs from \emph{trainable} parameters, interleaving the two kinds of rotation and uploading the same input into several of them~\cite{PerezSalinas2020datareuploading}; the \texttt{embedding\_idx} and \texttt{param\_idx} tags of \Cref{sec:compile-once} encode this split. A batch of inputs is passed as \texttt{embedding} with shape $(B_e, N_{\mathrm{emb}})$, $N_{\mathrm{emb}}$ being one more than the largest \texttt{embedding\_idx} in the circuit (\texttt{qc.n\_embedding}), with one \texttt{thetas} shared across the batch.

The XOR classifier of \Cref{lst:xor} embeds a two-dimensional input through two rotations driven by embedding inputs, entangles with $\mathrm{CNOT}$s, maps $\langle Z_0\rangle\in[-1,1]$ affinely to a class probability, and trains its parameters against a binary cross-entropy loss on $512$ points drawn uniformly from $[-\pi,\pi]^2$ and labeled $1$ where $x_1x_2<0$, in minibatches of $128$.

\begin{lstlisting}[caption={Nonlinear (XOR) classification with batched embedding inputs (condensed).},label={lst:xor}]
# Rotations driven by embedding inputs use embedding_idx; only the param_idx angles are trained.
qc = Circuit(n_qubits=2)
qc.rx(0, embedding_idx=0).rx(1, embedding_idx=1).cnot(0, 1)
qc.ry(0, param_idx=0).ry(1, param_idx=1).cnot(1, 0)
qc.ry(0, param_idx=2)
qc.compile(observables=[("Z", [0])], preset="gpu")

thetas = torch.nn.Parameter(0.1 * torch.randn(3, dtype=torch.float64, device="cuda"))
opt = torch.optim.Adam([thetas], lr=0.03)
for epoch in range(320):
    # X_batch: (B, 2) embedding inputs; y_batch: (B,) labels in {0, 1}
    expv = qc.expvals(thetas, embedding=X_batch)[:, 0]          # (B,)
    prob = torch.clamp((expv + 1.0) * 0.5, 1e-5, 1.0 - 1e-5)
    loss = torch.nn.functional.binary_cross_entropy(prob, y_batch)
    opt.zero_grad(set_to_none=True); loss.backward(); opt.step()
\end{lstlisting}

\begin{figure}[b]
    \centering
    \begin{subfigure}[b]{0.36\linewidth}
        \centering
        \includegraphics[width=\linewidth]{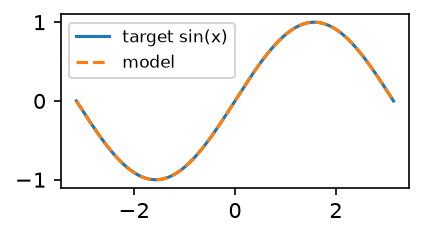}
        \caption{$y=\sin x$ regression}
    \end{subfigure}
    \hfill
    \begin{subfigure}[b]{0.31\linewidth}
        \centering
        \includegraphics[width=\linewidth]{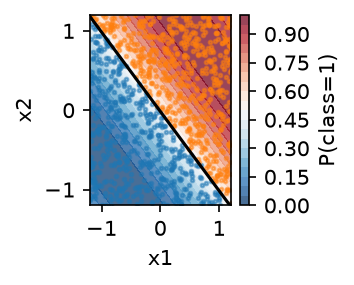}
        \caption{linear classification}
    \end{subfigure}
    \hfill
    \begin{subfigure}[b]{0.31\linewidth}
        \centering
        \includegraphics[width=\linewidth]{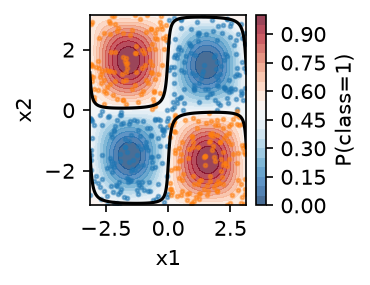}
        \caption{XOR classification}
    \end{subfigure}
    \caption{Trained models of the three embedding tasks of \Cref{sec:ex-embed}; \emph{(c)} is the model of \Cref{lst:xor}. In \emph{(b)} and \emph{(c)}: shading, the model's $P(\text{class}{=}1)$; points, the training set (class $0$ blue, class $1$ orange); black contours, the $P=0.5$ decision boundary.}
    \label{fig:embed}
\end{figure}

Two easier companion tasks, a single-qubit model fitting $y=\sin x$ and a linearly separable two-class problem, pass their inputs through \texttt{embedding} in the same way; \Cref{fig:embed} shows the three trained models, the regression at a mean squared error of $7.0\times10^{-19}$ and the two classifiers at $99.7\%$ and $94.9\%$ training accuracy.

\paragraph{Ablation of zero-filtering.}
We quantify what zero-filtering contributes within the engine by compiling the same program with the pass disabled (\texttt{zero\_filter=False}, which keeps every propagated term). The instance is the $16$-qubit, depth-$5$ member of an Erd\H{o}s--R\'enyi MaxCut-\qaoa\ family (edge probability $0.3$, \texttt{numpy.random.default\_rng(42)}): a Hadamard layer on all qubits, then per layer one $ZZ$ cost rotation per edge and one $X$ mixer rotation per qubit, each with its own angle. We compile it under the \texttt{gpu} preset in \texttt{float32} on an AMD Instinct MI300X and sweep $w_{\max}=4$--$7$ (\Cref{tab:zf-ablation}). The two variants return identical expectation values at every $w_{\max}$, and their gradients (\texttt{vjp} mode) agree to $\le3.6\times10^{-7}$, at single-precision round-off, an empirical confirmation that zero-filtering introduces no approximation error.

On the cost side, zero-filtering shrinks the evaluation working set by two to three orders of magnitude over this sweep, from $100\times$ at $w_{\max}=4$ to $1{,}075\times$ at $w_{\max}=7$, while the explicitly stored entries of the compiled step matrices fall only $2.2$--$3.8\times$. Each evaluation reads every stored entry of those matrices, and the measured wall-clock saving is closer to the reduction in stored entries than to the reduction in terms.

The batched forward pass improves by about $1.5\times$ ($1.47$--$1.65\times$) across the sweep, and steady-state memory by $1.4$--$1.5\times$ at $w_{\max}=5$--$7$ ($3.0\times$ at $w_{\max}=4$). The warm forward-plus-backward pass, taken after the backward pass's cached matrices exist, gains less: $1.1\times$ and $1.4\times$ at $w_{\max}=6$ and $7$, and at $w_{\max}=4$ and $5$ it is nearly independent of the term count, consistent with a fixed per-query overhead dominating there.

Zero-filtering also lowers peak memory: at $w_{\max}=7$ the unfiltered forward-plus-backward pass peaks at $119.8$\,GB of allocated GPU memory against $74.4$\,GB with zero-filtering, and the batched forward pass over $200$ parameter vectors at $156.0$\,GB against $126.5$\,GB. (Allocated memory is what PyTorch's allocator has handed to live tensors; the reserved memory of \Cref{sec:resource-scaling} adds the allocator's cached blocks, and the occupancy of \Cref{sec:benchmarks} is the device total read from the driver.)

The first forward-plus-backward pass, which builds the matrices the backward pass reads, is a one-time cost of $0.8$\,s at $w_{\max}=4$ to $161.1$\,s at $w_{\max}=7$, the whole gradient time for a workload that computes a single gradient. Disabling zero-filtering shortens compilation by $0.2$--$1.0$\,s, less than zero-filtering saves on that first pass alone ($0.35$--$58$\,s).

\begin{table}[t]
    \centering
    \small
    \begin{tabular}{clrrrrrr}
        \toprule
         & & & & & & \multicolumn{2}{c}{Fwd$+$bwd [s]} \\
        \cmidrule(lr){7-8}
        $w_{\max}$ & Zero-filtering & Terms & Step nonzeros & Steady-state [GB] & Eval$_{200}$ [s] & warm & first \\
        \midrule
        \multirow{2}{*}{4} & on  & $696$              & $4.85\times10^{5}$ & $0.06$ & $0.068$ & $0.067$ & $0.8$ \\
                           & off & $69{,}665$         & $1.82\times10^{6}$ & $0.18$ & $0.100$ & $0.068$ & $1.2$ \\
        \addlinespace
        \multirow{2}{*}{5} & on  & $2{,}516$          & $5.81\times10^{6}$ & $1.09$ & $0.769$ & $0.070$ & $5.6$ \\
                           & off & $584{,}910$        & $1.76\times10^{7}$ & $1.65$ & $1.266$ & $0.071$ & $8.4$ \\
        \addlinespace
        \multirow{2}{*}{6} & on  & $6{,}884$          & $5.17\times10^{7}$ & $7.61$ & $5.201$ & $0.085$ & $35.8$ \\
                           & off & $3{,}504{,}271$    & $1.26\times10^{8}$ & $10.47$ & $7.972$ & $0.095$ & $50.3$ \\
        \addlinespace
        \multirow{2}{*}{7} & on  & $14{,}892$         & $3.10\times10^{8}$ & $37.03$ & $22.857$ & $0.194$ & $161.1$ \\
                           & off & $16{,}011{,}510$   & $6.68\times10^{8}$ & $50.28$ & $33.520$ & $0.263$ & $219.3$ \\
        \bottomrule
    \end{tabular}
    \caption{Ablation of zero-filtering on the $16$-qubit, depth-$5$ Erd\H{o}s--R\'enyi MaxCut-\qaoa\ instance (AMD Instinct MI300X, \texttt{float32}; medians of five calls after one warm-up, except \emph{first}, which is timed once). Terms, evaluation working set; Step nonzeros, stored entries of the step matrices; Steady-state, GPU memory held between forward queries; Eval$_{200}$, batched forward pass over $200$ parameter vectors; Fwd$+$bwd, forward-plus-backward pass, \emph{warm} with the backward cache built and \emph{first} for the pass that builds it.}
    \label{tab:zf-ablation}
\end{table}

\paragraph{Cost of the two differentiation modes.}
\Cref{tab:diff-mode} sweeps $w_{\max}=4$--$7$ on the instance of \Cref{tab:zf-ablation}. The warm forward-plus-backward pass of \texttt{vjp} is one to nearly three orders of magnitude faster, a gap that widens with the working set, and its peak is smaller throughout, by $7\%$ at $w_{\max}=4$ and $11$--$12\%$ from $w_{\max}=5$ on. The \texttt{autograd} times coincide with the \emph{first} forward-plus-backward pass of \Cref{tab:zf-ablation} (within $1.1\%$ from $w_{\max}=5$ on and $8\%$ at $w_{\max}=4$, from the unrounded times), which indicates that the time gap is the cost of forming the sparse matrices that \texttt{vjp} caches and the traced path re-forms on every call; the smaller peak is consistent with \texttt{vjp} not retaining the autograd graph. The two gradients agree to $4.8\times10^{-7}$, single-precision round-off; \texttt{autograd} is kept as an independent cross-check.

\begin{table}[t]
    \centering
    \small
    \begin{tabular}{c r r r r r r}
        \toprule
        & \multicolumn{2}{c}{Fwd$+$bwd [s]} & \multicolumn{2}{c}{Peak [GB]} & & \\
        \cmidrule(lr){2-3} \cmidrule(lr){4-5}
        $w_{\max}$ & \texttt{vjp} & \texttt{autograd} & \texttt{vjp} & \texttt{autograd} & Memory saved & Speedup \\
        \midrule
        $4$ & $0.068$ & $0.76$ & $0.341$ & $0.368$ & $7\%$ & $11\times$ \\
        $5$ & $0.071$ & $5.58$ & $2.143$ & $2.413$ & $11\%$ & $78\times$ \\
        $6$ & $0.087$ & $35.8$ & $14.7$ & $16.8$ & $12\%$ & $412\times$ \\
        $7$ & $0.193$ & $161.0$ & $74.4$ & $84.4$ & $12\%$ & $832\times$ \\
        \bottomrule
    \end{tabular}
    \caption{The two differentiation modes on the instance of \Cref{tab:zf-ablation}, under its timing protocol. Fwd$+$bwd, one warm forward-plus-backward pass; Peak, its peak allocated GPU memory; Memory saved, one minus the ratio of the two peaks; Speedup, the \texttt{autograd}-to-\texttt{vjp} time ratio.}
    \label{tab:diff-mode}
\end{table}

\subsection{Interface, presets, and scope}
\label{sec:api}

This subsection documents the code architecture, the supported gate set, the cost of the execution presets, the quasi-probability sampler, and the reference helpers; \Cref{app:internals} catalogs the objects the interface compiles to.

\paragraph{Code architecture and backend implementation.}
\pado\ is a Python frontend over a C\texttt{++} extension. The frontend defines circuits and observables; during compilation the extension performs the operations applied at each gate (Pauli mask rewrites, branch generation under rotations, merging of duplicate terms, truncation, and step construction for the two single-qubit noise channels) and returns the index and value tensors of each step.

The extension finds identical strings by comparing their mask rows, merges them, and applies the truncation rules in the same pass, so branches that fail a truncation rule are dropped inside the extension call and never reach Python. Python assembles the returned tensors into the sparse step matrices of \Cref{sec:compile-once}, so evaluation and differentiation run as sparse tensor kernels on the preset's compute device inside PyTorch's autograd graph.

\paragraph{Supported gate set and scope.}
\label{sec:scope}
The engine has three kinds of primitive: Pauli rotations, a small set of Clifford gates, and two single-qubit noise channels.

\begin{itemize}[leftmargin=1.2em,itemsep=0.3em]
    \item \textbf{Pauli rotations.} \texttt{PauliRotation} accepts an \emph{arbitrary} Pauli string generator $G$ and realizes $e^{-i\theta G/2}$, so any single- or multi-qubit Pauli rotation (e.g.\ \texttt{X}, \texttt{ZZ}, \texttt{XYZ}) is a primitive. These are the only branching unitary elements and hence the main source of term growth; amplitude damping also branches, but only by splitting a $Z$ on the damped qubit into $Z$ and the identity (\Cref{sec:noise}). A rotation at a Clifford angle, such as $R_{ZZ}(-\pi/2)$, is still a rotation to the engine; its vanishing branch is removed only by coefficient truncation (\Cref{sec:ex-kicked-ising}).
    \item \textbf{Clifford primitives.} The engine natively conjugates H, S, X, CNOT, and CZ; other Cliffords (Y, Z, SX, SWAP, and the like) must be written as short products of these, which adds no term growth since a Clifford maps each Pauli string to one Pauli string.
    \item \textbf{Noise channels.} \texttt{DepolarizingNoise} and \texttt{AmplitudeDampingNoise} represent the two single-qubit channels of \Cref{sec:noise}.
\end{itemize}

Other gates (arbitrary non-Clifford two-qubit unitaries, operations with many controls) must be decomposed into these primitives; such constructions can branch heavily.

\paragraph{Preset comparison.}
\Cref{tab:preset-comparison} measures what each preset costs on the $16$-qubit, depth-$5$ instance of \Cref{tab:zf-ablation} at $w_{\max}=4$--$6$, compiled once per preset at the shared \texttt{float64} default on the MI300X and timed as the median of twenty calls (five for the batched pass) after one discarded warm-up. The three presets' values and gradients agree to $5.6\times10^{-16}$. Moving storage and compute to the GPU (the \texttt{cpu} against the \texttt{gpu} rows) speeds up the batched forward pass (Eval$_{200}$) by $25$--$30\times$ across the sweep, the single evaluation (Eval) by $1.8$--$16\times$, and the warm forward-plus-backward pass (Fwd$+$bwd) by $1.5$--$14\times$, the latter two growing with the evaluation working set. The \texttt{hybrid} preset is close to \texttt{gpu} on Eval$_{200}$ and, from $w_{\max}=5$ on, between \texttt{cpu} and \texttt{gpu} on Eval and Fwd$+$bwd.

\begin{table}[t]
    \centering
    \small
    \begin{tabular}{c l r r r r}
        \toprule
        $w_{\max}$ & Preset & Compile [s] & Eval [ms] & Eval$_{200}$ [s] & Fwd$+$bwd [ms] \\
        \midrule
        \multirow{3}{*}{4} & \texttt{cpu}    & $1.32$ & $40.2$ & $2.25$ & $104$ \\
                           & \texttt{hybrid} & $0.82$ & $43.5$ & $0.11$ & $114$ \\
                           & \texttt{gpu}    & $0.62$ & $22.8$ & $0.089$ & $69.2$ \\
        \addlinespace
        \multirow{3}{*}{5} & \texttt{cpu}    & $3.32$ & $117.1$ & $25.8$ & $327$ \\
                           & \texttt{hybrid} & $2.11$ & $64.2$ & $0.91$ & $156$ \\
                           & \texttt{gpu}    & $3.58$ & $23.5$ & $0.87$ & $72.4$ \\
        \addlinespace
        \multirow{3}{*}{6} & \texttt{cpu}    & $30.69$ & $611.6$ & $150.6$ & $1{,}302$ \\
                           & \texttt{hybrid} & $10.01$ & $127.4$ & $5.91$ & $280$ \\
                           & \texttt{gpu}    & $5.51$ & $37.6$ & $5.82$ & $96.1$ \\
        \bottomrule
    \end{tabular}
    \caption{Preset comparison on the instance of \Cref{tab:zf-ablation} (\texttt{float64}, MI300X). Eval, one forward pass on one parameter vector; Eval$_{200}$, one batched pass over $200$; Fwd$+$bwd, one warm forward-plus-backward pass in \texttt{vjp} mode; \texttt{cpu} at PyTorch's default of $20$ threads.}
    \label{tab:preset-comparison}
\end{table}

\paragraph{Quasi-probability reconstruction.}
Expectation values of $Z$-strings also determine the circuit's output distribution: the Born rule probabilities are the Walsh--Hadamard transform of the correlators $\langle Z_S\rangle_\theta$ over all qubit subsets $S$ ($Z_S$ the product of $Z$ on the qubits in $S$)~\cite{herrerogonzalez2025bornultimatumconditionsclassical}, and keeping only subsets of order at most $k$ gives the $k$-order truncation $q^{(k)}(x)=2^{-n}\sum_{|S|\le k}(-1)^{\sum_{i\in S}x_i}\langle Z_S\rangle_\theta$, a signed quasi-probability that sums to one but may be negative~\cite{herrerogonzalez2025bornultimatumconditionsclassical}; truncated Fourier series of output distributions and their marginals are computed as quasi-probability distributions in~\cite{cirstoiu2024fourieranalysisframeworkapproximate}; the number of samples needed to estimate probabilities from it grows with its $\ell_1$ norm~\cite{PhysRevLett.115.070501}. Its cost is the $\sum_{m=1}^{k}\binom{n}{m}$ retained correlators, which the compiled program supplies directly.

\path{build_quasi_sampler} implements this reconstruction: it compiles a sampler from the circuit and the $Z$-string subsets to track (\texttt{z\_combos}, a list of qubit subsets), accepts the presets of \Cref{tab:presets}, and provides \texttt{compute\_moments} for the correlators and \texttt{quasi\_prob} for the reconstructed distribution (\Cref{lst:quasi}). For training, \path{quasi_prob_from_moments} reconstructs $q^{(k)}(x)$ from cached moments across many parameter and bitstring batches. At full order the reconstruction is exact: on a $10$-qubit test circuit (a Hadamard layer, then two layers of per-qubit $Z$ and $X$ rotations and a ring of $ZZ$ rotations) compiled without max-weight truncation, it agrees with the exact PennyLane probabilities over all $2^{10}$ bitstrings to a total variation distance of $5\times10^{-15}$, and to $0.02$ when truncated at order $k=8$ of $10$.

\begin{lstlisting}[caption={Moment computation and quasi-probability reconstruction.},label={lst:quasi}]
import torch
from itertools import combinations
from padopauli import Circuit, build_quasi_sampler

n_qubits, max_order = 4, 4
qc = Circuit(n_qubits=n_qubits)
qc.h(0).cnot(0, 1).rzz(1, 2, param_idx=0).rxx(2, 3, param_idx=1)

z_combos = [list(c) for k in range(1, max_order + 1)
            for c in combinations(range(n_qubits), k)]
sampler = build_quasi_sampler(n_qubits=n_qubits, circuit=qc.gates,
                              z_combos=z_combos, max_order=max_order, preset="gpu")

# x_batch: (B, n_qubits) uint8 bitstrings, the x at which q(x) is wanted.
thetas = torch.tensor([0.3, -0.4], dtype=torch.float64)
moments = sampler.compute_moments(thetas)                 # Z-correlator moments
q = sampler.quasi_prob(x_batch, thetas, order=max_order)  # reconstruction in one call
\end{lstlisting}

\paragraph{PennyLane reference helpers.}
\label{sec:ex-ref}
For small systems the engine includes reference helpers that call PennyLane, an optional dependency installed through the \texttt{padopauli[reference]} extra. A compiled program's \path{expvals_reference} method returns exact expectation values; the top-level helpers \path{pennylane_expvals}, \path{pennylane_sample}, and \path{pennylane_probs} return exact expectation values, samples, and probabilities from a gate list. All are capped at $n\le 20$ by default (\texttt{max\_qubits} sets the cap) and let a user check correctness before a large run.

\section{Truncation Accuracy and Resource Cost}
\label{sec:trunc-cost}

Apart from floating-point precision, the only approximation \pado\ introduces is the discarding of Pauli terms, so its error is bounded by \Cref{eq:trunc-l1}. The accuracy results below are measured against exact baselines.

\subsection{Accuracy against exact baselines: convergence and exact checks}
\label{sec:trunc-accuracy}
The worst-case bound can be loose, so we check accuracy empirically in two ways. For small systems, the PennyLane~\cite{bergholm2022pennylaneautomaticdifferentiationhybrid} reference helpers of \Cref{sec:ex-ref} supply an exact statevector baseline, against which we measure the true error of a truncation directly. For larger systems, where no exact baseline is available, we monitor \emph{apparent convergence} \cite{6vd7-l9bn,gharibyan2025practicalguideusingpauli}: we recompute the expectation values under progressively looser truncation (larger $w_{\max}$, smaller $\epsilon$) and accept the operating point once they stabilize. That is a diagnostic rather than a guarantee: loosening the truncation does not always improve accuracy, and some instances that reproduce hardware results are not converged \cite{gharibyan2025practicalguideusingpauli}. \Cref{sec:ex-kicked-ising} runs it at $127$ qubits.

\begin{figure}[t]
    \centering
    \begin{subfigure}[b]{0.49\linewidth}
        \centering
        \includegraphics[width=\linewidth]{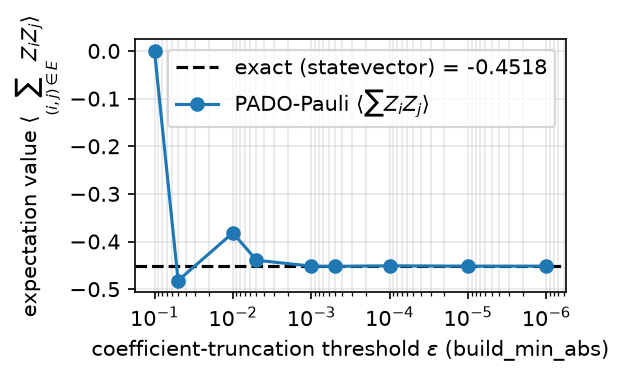}
        \caption{expectation value vs.\ $\epsilon$}
    \end{subfigure}
    \hfill
    \begin{subfigure}[b]{0.49\linewidth}
        \centering
        \includegraphics[width=\linewidth]{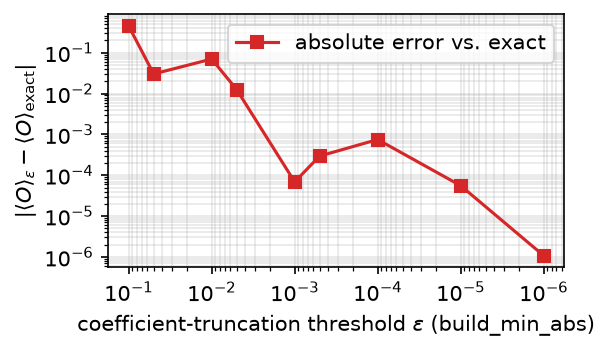}
        \caption{absolute error vs.\ $\epsilon$}
    \end{subfigure}
    \caption{Convergence of coefficient truncation on a fixed $16$-qubit, depth-$3$ Erd\H{o}s--R\'enyi MaxCut-\qaoa\ instance with max-weight truncation disabled. \emph{(a)} truncated expectation value against the exact statevector value (dashed); \emph{(b)} absolute error. In both panels the $\epsilon$ axis is logarithmic and reversed, running from $10^{-1}$ at the left to $10^{-6}$ at the right.}
    \label{fig:minabs-conv}
\end{figure}

\paragraph{Convergence of coefficient truncation to the exact value.}
The first check uses a fixed instance (\Cref{fig:minabs-conv}): a $16$-qubit, depth-$3$ member of the Erd\H{o}s--R\'enyi MaxCut-\qaoa\ family of \Cref{tab:zf-ablation} (edge probability $0.3$, seed $42$) at the fixed parameter vector $\theta_k=0.2\sin k$, against an exact \texttt{default.qubit} statevector baseline. All accuracy studies here run the compiled program and its baselines in \texttt{float64}, so the reported errors are truncation errors, and a retained count is the size of the zero-filtered evaluation working set. With max-weight truncation disabled ($w_{\max}=n$, the largest reachable weight), coefficient truncation is the only approximation. Over a sweep of its threshold from $\epsilon=10^{-1}$ to $10^{-6}$, the absolute error on $\langle\sum_{(i,j)\in E}Z_iZ_j\rangle$ ($E$ the graph's edge set) falls from $0.45$ at $\epsilon=10^{-1}$, where every term is discarded, to $1.1\times10^{-6}$ at $\epsilon=10^{-6}$ on $152$ retained Pauli strings. The convergence is not strictly monotone, a known feature of coefficient truncation \cite{gharibyan2025practicalguideusingpauli} consistent with \Cref{eq:trunc-l1} being an upper bound.

\begin{figure}[t]
    \centering
    \includegraphics[width=0.74\linewidth]{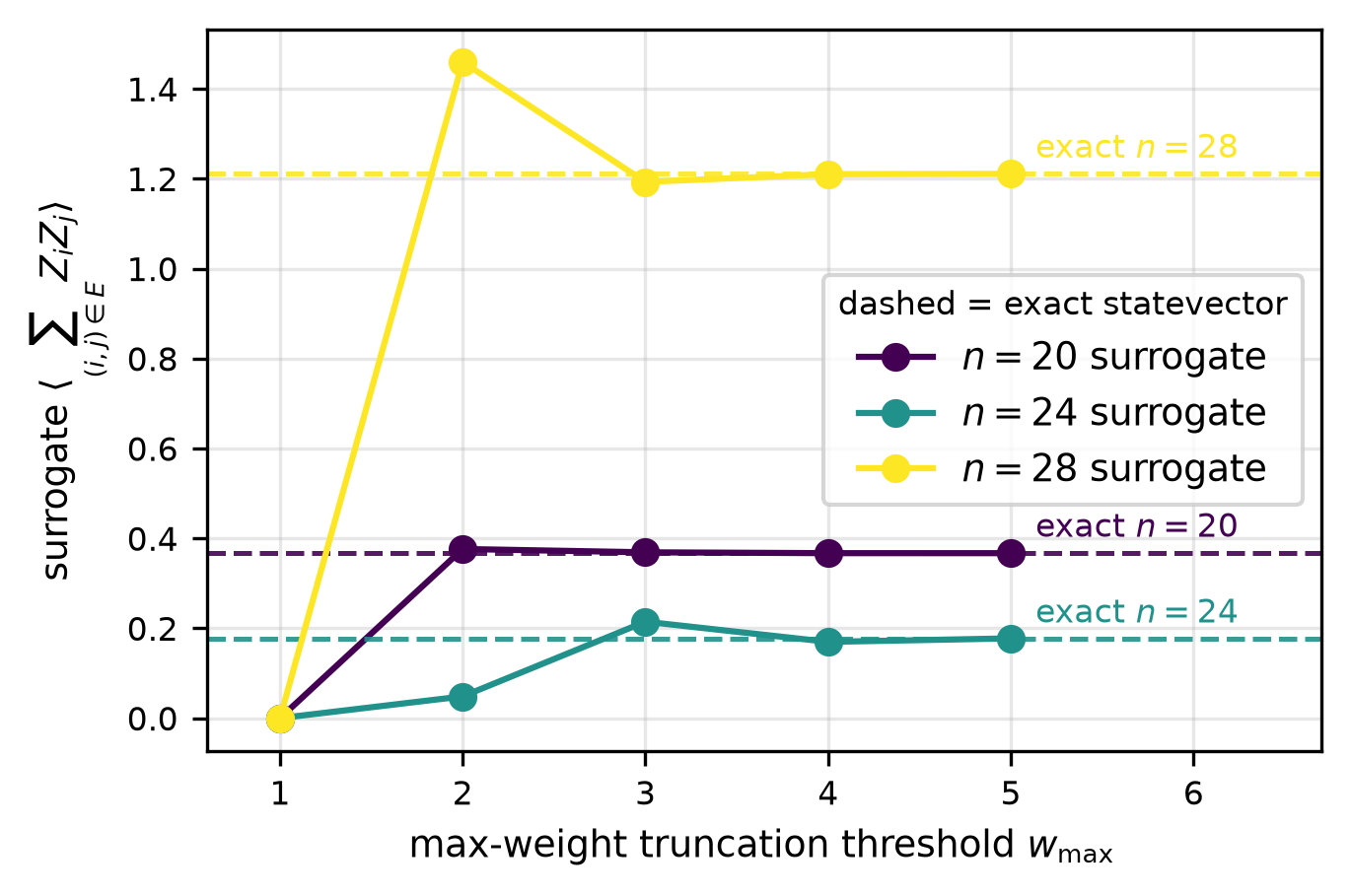}
    \caption{Convergence to the exact value across system sizes on the depth-$3$ Erd\H{o}s--R\'enyi MaxCut-\qaoa\ family (seed $42$, $\theta_k=0.2\sin k$) at $n=20$, $24$, $28$. Solid, the truncated $\langle\sum_{(i,j)\in E}Z_iZ_j\rangle$ as $w_{\max}$ is raised from $1$ to $5$; dashed, the exact \texttt{lightning.amdgpu} values in matching color.}
    \label{fig:accuracy-scale}
\end{figure}

\paragraph{Convergence to the exact value across system sizes.}
\Cref{fig:accuracy-scale} applies the same check to the max-weight threshold: $w_{\max}=1$ to $5$ on the same family (depth $3$, edge probability $0.3$, seed $42$) at the same $\theta_k=0.2\sin k$, now at $n=20$, $24$, and $28$ qubits, against exact statevector baselines from PennyLane's \texttt{lightning.amdgpu} (dashed lines). At every size the truncated $\langle\sum_{(i,j)\in E}Z_iZ_j\rangle$ converges onto its exact value, though not monotonically: the curves overshoot and cross before converging (most visibly $n=28$ at $w_{\max}=2$ and $n=24$ at $w_{\max}=3$), so successive changes shrink only from $w_{\max}=3$ onward. By $w_{\max}=5$ the absolute error is at most $9.74\times10^{-4}$, against exact values $0.367$, $0.176$, and $1.211$ at $n=20$, $24$, and $28$; the largest error is at $n=24$, where the observable is also smallest.

Two further graph draws (seeds $7$ and $123$) show the same convergence: over the nine size--seed pairs the $w_{\max}=5$ error ranges from $3.2\times10^{-7}$ to $9.74\times10^{-4}$, and the numbers of retained terms at $w_{\max}=5$ ($5.8\times10^{3}$ to $2.4\times10^{4}$ Pauli strings) agree across seeds to within $8\%$ at every size. These errors are measured at the small angles $\theta_k=0.2\sin k$; on a different instance at trained angles, the same $w_{\max}=5$ leaves an error of $0.036$ on a $30$-edge objective (\Cref{sec:ex-maxcut}).

\begin{figure}[t]
    \centering
    \begin{subfigure}[b]{0.48\linewidth}
        \centering
        \includegraphics[width=\linewidth]{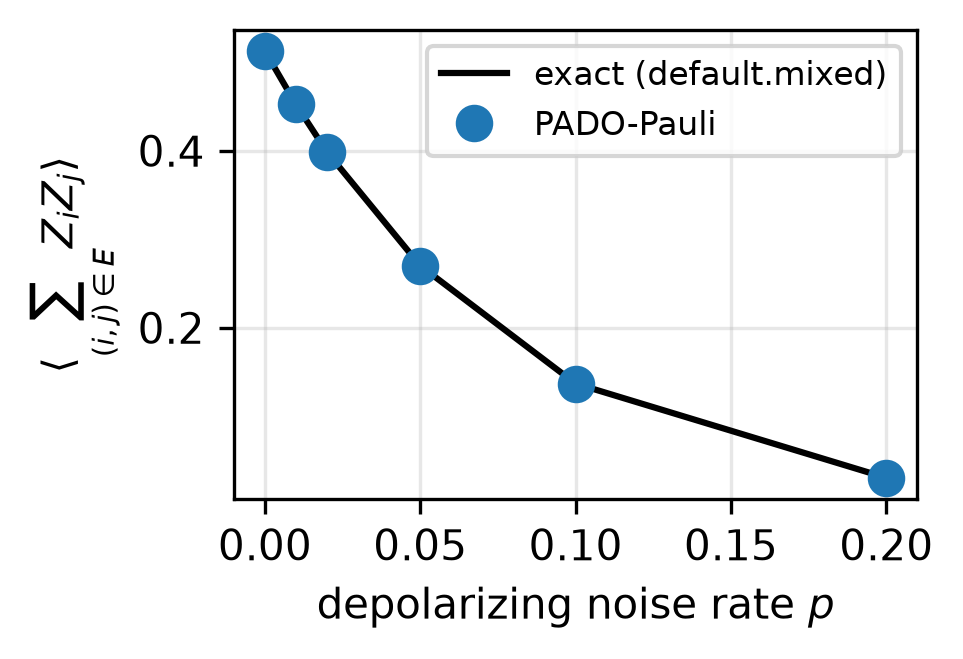}
        \caption{depolarizing: correctness vs.\ $p$}
        \label{fig:noise-validation}
    \end{subfigure}
    \hfill
    \begin{subfigure}[b]{0.48\linewidth}
        \centering
        \includegraphics[width=\linewidth]{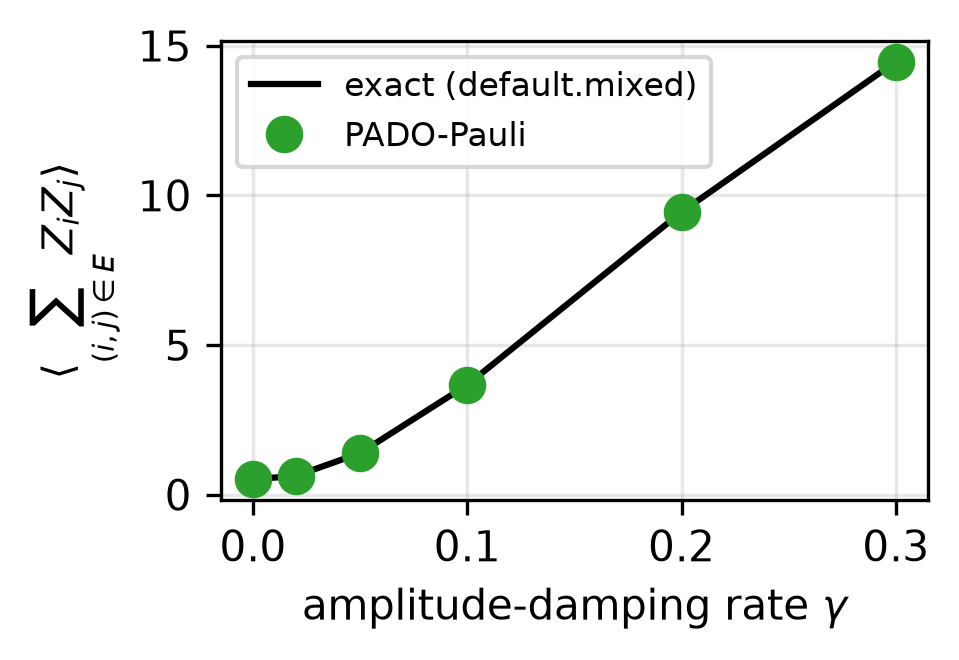}
        \caption{amplitude damping: correctness vs.\ $\gamma$}
        \label{fig:noise-ampdamp}
    \end{subfigure}\\[0.6em]
    \begin{subfigure}[b]{0.48\linewidth}
        \centering
        \includegraphics[width=\linewidth]{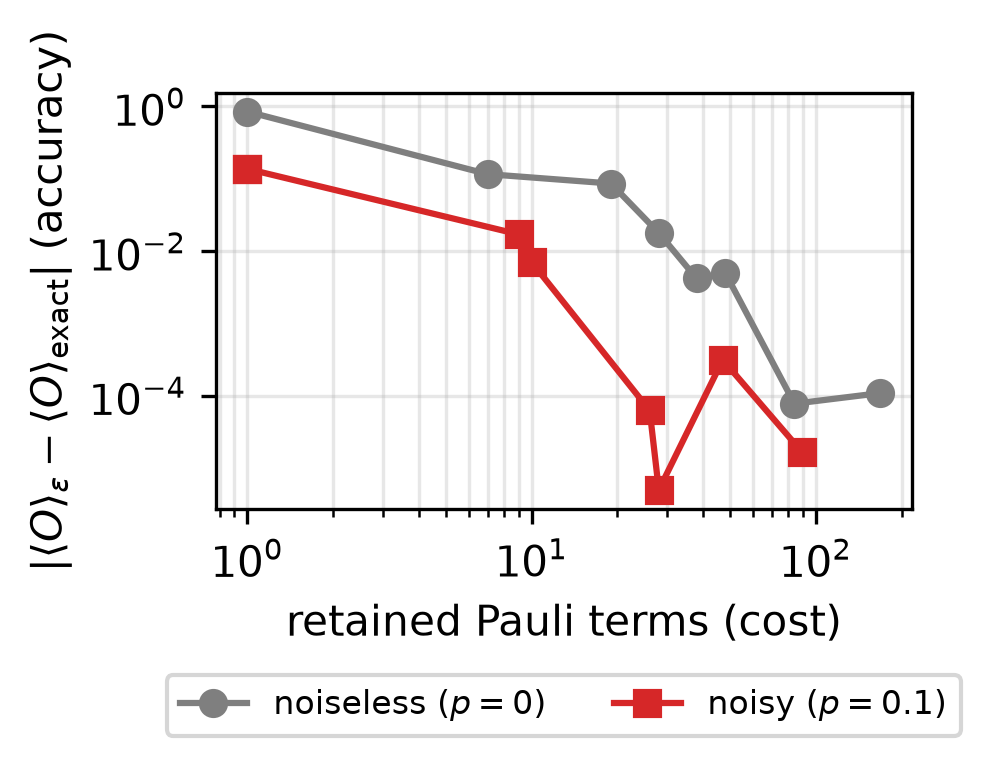}
        \caption{accuracy--cost, noisy vs.\ noiseless}
        \label{fig:noise-truncation}
    \end{subfigure}
    \caption{Noisy circuit studies on a $10$-qubit, depth-$5$ Erd\H{o}s--R\'enyi MaxCut-\qaoa\ instance against PennyLane's exact \texttt{default.mixed} simulator. \emph{(a)} depolarizing noise of rate $p$ and \emph{(b)} amplitude damping of rate $\gamma$ after every layer, \pado\ as markers, the simulator as the line; \emph{(c)} absolute error against retained Pauli strings under a coefficient truncation sweep, noiseless (gray) and noisy at $p=0.1$ (red); the leftmost red marker, drawn at one retained string, is the empty program at $\epsilon\ge0.05$.}
    \label{fig:noise-studies}
\end{figure}

\paragraph{Noisy circuits: correctness and the truncation benefit of noise.}
Two things are checked: that \pado\ propagates noise channels correctly, and that noise lowers the truncation error at a fixed term budget, as \Cref{sec:noise} predicts. Both use a $10$-qubit, depth-$5$ Erd\H{o}s--R\'enyi MaxCut-\qaoa\ instance (edge probability $0.5$, seed $42$) at $\theta_k=0.2\sin k$, with a single-qubit channel (depolarizing of per-axis rate $p/3$, or amplitude damping of rate $\gamma$) on every qubit after each layer, against PennyLane's \texttt{default.mixed} density matrix simulator (\Cref{fig:noise-studies}). The truncation benefit is analyzed in \Cref{sec:frontier}.

\paragraph{Correctness against an exact density matrix simulator.}
Sweeping the depolarizing rate from $p=0$ to $0.2$ at $\epsilon=10^{-6}$, the truncated $\langle\sum_{(i,j)\in E}Z_iZ_j\rangle$ agrees with the density matrix value to within $2.52\times10^{-5}$ over the whole sweep while the observable decays from $0.513$ to $0.031$ (\Cref{fig:noise-validation}), confirming that the diagonal Pauli transfer action of \Cref{eq:depol-ptm} is implemented correctly. For amplitude damping, a non-unital channel propagated through its unital Heisenberg dual (\Cref{sec:noise}), the agreement is within $3.72\times10^{-5}$ over the whole sweep $\gamma\in[0,0.3]$ (\Cref{fig:noise-ampdamp}); there the observable rises from $0.513$ to $14.5$ as each qubit relaxes toward $\ket{0}$.

\subsection{Growth of Pauli terms and resource scaling}
\label{sec:resource-scaling}

This subsection measures how large the compiled program becomes after propagation, truncation, and zero-filtering, and what that costs in runtime and memory.

\paragraph{Max-weight sweep at fixed system size.}
The first study holds the circuit fixed and varies only $w_{\max}$, from $2$ to $5$. The circuit is a $15$-qubit, depth-$5$ instance of the Erd\H{o}s--R\'enyi MaxCut-\qaoa\ family of \Cref{tab:zf-ablation} (edge probability $0.3$, seed $42$) with cost observable $\sum_{(i,j)\in E} Z_iZ_j$; max-weight truncation is structural, so the term counts do not depend on the angles. The numbers of non-identity Pauli strings of weight at most $w_{\max}$, upper bounds on the term counts reported below, are
\begin{equation}
    \sum_{w=1}^{w_{\max}} \binom{n}{w} 3^w
    \qquad \text{and} \qquad
    \sum_{w=1}^{w_{\max}} \binom{n}{w},
\end{equation}
for the full Pauli basis and its diagonal $Z$-type sector, respectively.

Raising $w_{\max}$ from $2$ to $5$ gives unique propagated bases of $135$, $3{,}247$, $51{,}794$, $401{,}871$ terms (factors $24$, $16$, $7.8$), while the zero-filtered sector grows more slowly, $15$, $120$, $575$, $1{,}940$ (factors $8$, $4.8$, $3.4$): exactly the complete diagonal $Z$-type sector one weight level below the threshold, $\sum_{w=1}^{w_{\max}-1}\binom{n}{w}$.

\begin{figure}[t]
    \centering
    \begin{subfigure}[b]{0.49\linewidth}
        \centering
        \includegraphics[width=\linewidth]{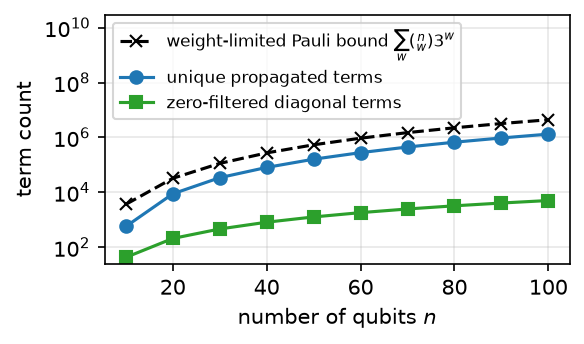}
        \caption{term counts vs.\ the upper bound}
        \label{fig:terms-vs-qubits}
    \end{subfigure}
    \hfill
    \begin{subfigure}[b]{0.49\linewidth}
        \centering
        \includegraphics[width=\linewidth]{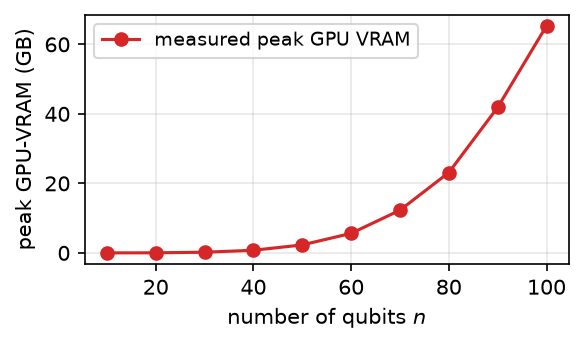}
        \caption{peak reserved GPU memory during compilation}
        \label{fig:vram-vs-qubits}
    \end{subfigure}
    \caption{The qubit sweep at $w_{\max}=3$ from $10$ to $100$ qubits (depth-$5$ Erd\H{o}s--R\'enyi MaxCut-\qaoa, edge probability $0.3$) under the \texttt{gpu} preset: \emph{(a)} propagated and zero-filtered term counts against the combinatorial upper bound, \emph{(b)} peak reserved GPU memory during their compilation, with coefficients in \texttt{float32} on an AMD Instinct MI300X.}
    \label{fig:scaling}
\end{figure}

\paragraph{Scaling with qubit count at \texorpdfstring{$w_{\max}=3$}{wmax=3}.}
The second study fixes $w_{\max}=3$ and raises the qubit count from $10$ to $100$ in steps of $10$ on the same family (depth $5$, edge probability $0.3$, one graph per size, seed $42$). The threshold $w_{\max}=3$ is a cost setting, not an accuracy one: on the depth-$3$ family of \Cref{sec:trunc-accuracy} its worst error over the nine size--seed pairs is $3.8\times10^{-2}$, against $9.74\times10^{-4}$ at $w_{\max}=5$. The number of unique propagated terms rises from $574$ at $10$ qubits to $1{,}320{,}934$ at $100$, while the zero-filtered diagonal count grows from $43$ to $5{,}050$ (\Cref{fig:terms-vs-qubits}). From $20$ qubits on, this count equals $n+\binom{n}{2}$, the size of the full diagonal sector of weight at most $2$, since the dense graph makes every single- and two-qubit $Z$ string reachable.

The ratio of propagated terms to the upper bound $\sum_{w=1}^{w_{\max}}\binom{n}{w}3^w$ stabilizes near $0.30$ for the larger systems, while the ratio of zero-filtered to propagated terms falls monotonically from $0.075$ at $10$ qubits to $0.0038$ at $100$.

\paragraph{Runtime and memory implications.}
The same sweep gives the cost of this term growth (\Cref{fig:scaling}). Under the \texttt{gpu} preset in \texttt{float32} on an AMD Instinct MI300X, the propagation stage of compilation takes $0.98$\,s at $10$ qubits, $6.8$\,s at $60$, and $38.4$\,s at $100$; including the zero-filtering pass, the total compile time is $1.55$, $10.3$, and $47.9$\,s (medians of five compilations). On this sweep zero-filtering costs no more than the propagation it follows.

Under the \texttt{gpu} preset the compiled program and its construction intermediates are all held in GPU memory. The peak reserved GPU memory during compilation rises from $0.002$\,GB at $10$ qubits to $65.3$\,GB at $100$ (\Cref{fig:vram-vs-qubits}) and is set by the construction intermediates.

The times and memory above are the one-time compilation cost. After compilation, we evaluate each program on a single parameter vector and on a batch of $200$ and report, for each, the median of five timed calls after a discarded warm-up.

A single evaluation costs $8.6$\,ms at $10$ qubits and $822$\,ms at $100$, a factor of $95$. Over the sweep the qubit count grows $10\times$, the number of compiled steps (one per gate) $75\times$, and the entries stored across those steps $3{,}176\times$. The evaluation time tracks the step count: the time per step is $78$\,$\mu$s at $10$ qubits and $100$\,$\mu$s at $100$, a factor of $1.3$, while the time per stored nonzero falls from $2.3$\,$\mu$s to $68$\,ns, a factor of $33$.

A nearly constant cost per step holds only at small $w_{\max}$: in the $w_{\max}$ sweep of \Cref{tab:zf-ablation} the step list is fixed at $266$ steps, yet a single evaluation, timed in that \texttt{float32} run, grows from $22.0$ to $84.3$\,ms ($83$ to $317$\,$\mu$s per step) as the stored entries grow by $638\times$.

Batching gains most where the fixed cost of a call, the part independent of the number of parameter vectors, dominates. At $10$ qubits a batch of $200$ costs $9.3$\,ms, $46$\,$\mu$s per vector against $8.6$\,ms alone, a factor of $186$: almost the whole time of a single call is fixed cost. At $100$ qubits the batch costs $81.3$\,s, $406$\,ms per vector against $822$\,ms alone, a factor of $2.0$, consistent with each step matrix being read once per block of $200$ vectors.

In units of the engine's own evaluation time, the $47.9$\,s compilation at $100$ qubits equals about $60$ single evaluations or about $120$ batched parameter vectors; at $50$ qubits ($7.6$\,s compilation, $198$\,ms per evaluation) about $40$. Beyond that many evaluations, evaluation time dominates, which is the use the engine is designed for: the training runs of \Cref{sec:examples} that query the compiled program take $200$ to $500$ optimizer steps. \Cref{sec:benchmarks} gives the number of queries after which the compilation is amortized relative to another engine's per-query cost.

\subsection{Accuracy--cost frontier: max-weight vs.\ coefficient truncation}
\label{sec:frontier}
\begin{figure}[t]
    \centering
    \includegraphics[width=0.74\linewidth]{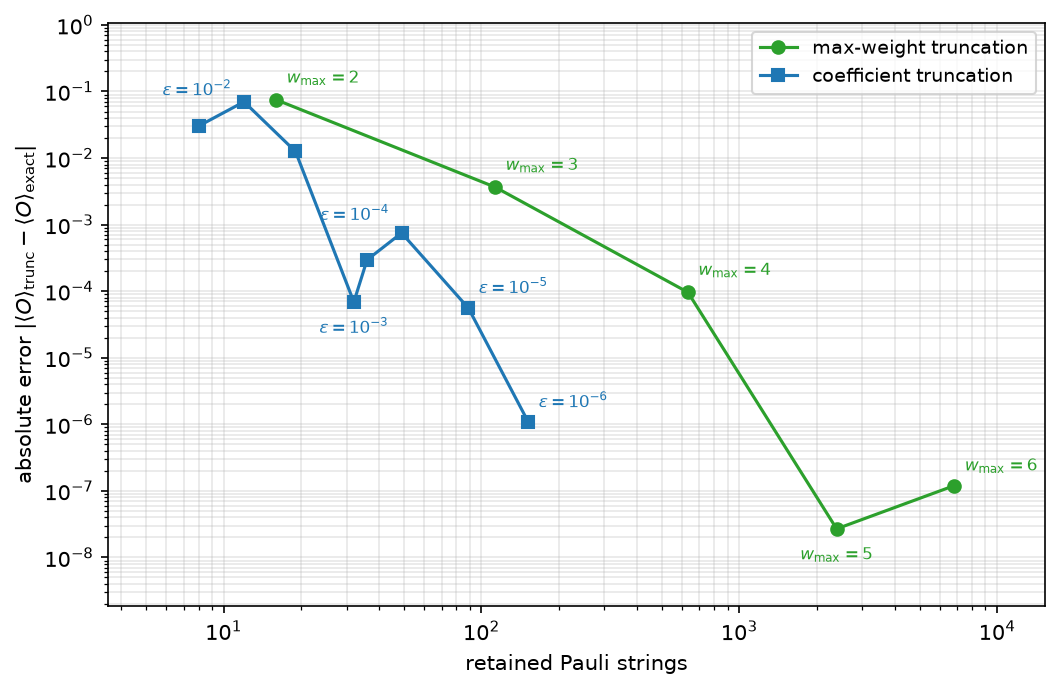}
    \caption{Accuracy--cost frontier on the fixed $16$-qubit, depth-$3$ Erd\H{o}s--R\'enyi MaxCut-\qaoa\ instance: absolute error against retained Pauli strings under a max-weight (green) and a coefficient (blue) sweep; each green marker is annotated with its $w_{\max}$, and each blue marker at a decade value of $\epsilon$ with that $\epsilon$. The coarsest setting of each sweep ($w_{\max}=1$, $\epsilon=10^{-1}$) retains no terms and is omitted.}
    \label{fig:accuracy-frontier}
\end{figure}

\Cref{fig:accuracy-frontier} compares the two truncation strategies on the $16$-qubit, depth-$3$ instance of \Cref{fig:minabs-conv}, plotting the true error against the number of retained Pauli strings as each threshold is swept: $w_{\max}=1$ to $6$ and $\epsilon=10^{-1}$ to $10^{-6}$. On this instance max-weight truncation converges smoothly and almost monotonically, to $2.7\times10^{-8}$ at $w_{\max}=5$ ($1.2\times10^{-7}$ at $6$); coefficient truncation, keeping the terms of largest magnitude irrespective of weight, reaches a given error with fewer terms, but not monotonically, as noted above. Coefficient truncation reaches $1.1\times10^{-6}$ at $152$ retained terms; max-weight truncation reaches $3.7\times10^{-3}$ at $113$ terms ($w_{\max}=3$) and $9.6\times10^{-5}$ at $632$ ($w_{\max}=4$), and falls below $1.1\times10^{-6}$ only at $2{,}385$ ($w_{\max}=5$).

Seeds $7$ and $123$ preserve this ordering at matched budgets above about $100$ retained strings (seed $7$: $2.0\times10^{-4}$ at $107$ against $6.6\times10^{-3}$ at $119$; seed $123$: $2.6\times10^{-5}$ at $160$ against $2.2\times10^{-3}$ at $132$), with the max-weight error $33$--$84$ times the coefficient truncation error. Coefficient truncation is therefore the better choice when the parameters are fixed and a small program is the priority, and max-weight truncation the safer default for training, where a term discarded at the build point receives no gradient.

\paragraph{Effect of noise on truncation cost.}
\Cref{fig:noise-truncation} repeats the coefficient truncation sweep ($\epsilon$ from $10^{-1}$ to $10^{-5}$) on the $10$-qubit instance of \Cref{fig:noise-validation} at $p=0$ and $p=0.1$, plotting the true error against retained Pauli strings. The noisy curve lies below and to the left of the noiseless one: an absolute error below $2\times10^{-3}$ first needs $84$ retained terms without noise but $26$ at $p=0.1$, about a threefold reduction in program size. As \Cref{sec:noise} predicts, each noisy layer multiplies coefficients by factors below one, so the noisy program retains fewer strings at every $\epsilon$ and a smaller set suffices.

\section{Performance Benchmarks against Other Engines}
\label{sec:benchmarks}

This section compares \pado\ with three Pauli propagation engines on a common workload. These are cuPauliProp, the GPU engine of the NVIDIA cuQuantum SDK; Qiskit's CPU pauli-prop path \cite{pauli-prop}, which truncates by a coefficient cutoff and a term cap ($10^{-14}$ and $10^{8}$ in every reported case); and PauliPropagation.jl \cite{6vd7-l9bn}, run on the CPU (in version~0.8.1 its CUDA extension covers only Pauli rotations and its README describes it as not yet performant \cite{paulipropagation_jl_repo}). PauliPropagation.jl is timed on two paths: the conventional path re-propagates on every call and returns the gradient through a paired forward and backward sweep (\texttt{rewindgradient} \cite{paulipropagation_jl_repo}); the experimental surrogate path builds the path graph once and re-evaluates it. \Cref{sec:bench-stress} additionally compares against exact statevector simulation.

The engines differ in three capabilities, which the timings reflect: GPU execution, a native backward pass, and batched embedding inputs. \pado\ is the only engine with all three. cuPauliProp has the first two; its truncation has a coefficient cutoff and a weight cutoff, with no zero-filtering~\cite{cuquantum_cupauliprop}. Qiskit pauli-prop has none of the three. PauliPropagation.jl has a native backward pass, no batched path, and GPU execution only through that extension; it also deletes off-diagonal strings through \texttt{zerofilter!} and reuses a compiled object on its surrogate path. Within \pado, GPU execution alone gives $1.8$--$16\times$ on single evaluations and $25$--$30\times$ on the batched pass over the \texttt{cpu} preset (\Cref{tab:preset-comparison}); the gaps below mix that effect with other implementation differences.

\subsection{Setup and numerical agreement}
\label{sec:bench-setup}

\paragraph{Workload.}
The benchmark circuit is a depth-$3$ multi-angle \qaoa\ ansatz \cite{Herrman2022} for an Ising model on a 2D grid with nearest-neighbor (NN) and next-nearest-neighbor (NNN) couplings, starting from the uniform superposition $H^{\otimes n}\ket{0}^{\otimes n}$. Each layer applies one $Z_uZ_v$ rotation per coupling, one $Z_q$ rotation per qubit, and one $X_q$ mixer rotation per qubit; every rotation has its own angle, drawn from $\mathcal{N}(0,0.3^2)$. The observable is the Ising energy $\sum_{(u,v)\in E}J_{uv}Z_uZ_v+\sum_q h_qZ_q$ with $J_{uv}\sim\mathcal{N}(0,1)$ (NN), $J_{uv}\sim\mathcal{N}(0,0.3^2)$ (NNN), and $h_q\sim\mathcal{N}(0,0.1^2)$. Scaling instances range over $2\times2$ ($4$q, $6$ $ZZ$ couplings), $2\times3$ ($6$q), $3\times3$ ($9$q), $2\times5$ ($10$q), and $3\times4$ ($12$q, $29$ $ZZ$ couplings); the truncation sweep uses a fixed $16$-qubit $4\times4$ grid.

\paragraph{Hardware and timing protocol.}
The orchestrator generates each circuit once as an explicit gate list, so every engine propagates the identical circuit; it runs each engine in its own subprocess. All timings and memory readings of the engine comparison were recorded on the NVIDIA A100 of \Cref{tab:platforms}; the stress test of \Cref{sec:bench-stress} ran on the AMD Instinct MI300X. Each timing in this subsection and the next is the mean of the last four of five repetitions, the first being discarded as warm-up. The CPU engines ran at their default thread settings. For the two GPU engines each timed forward pass is the forward half of a forward-plus-backward pass (\pado\ in \texttt{vjp} mode, the cuPauliProp driver keeping every per-gate expansion for the backward pass); the CPU engines are timed forward only. Each engine's run of a case, all five repetitions included, was limited to $900$\,s on the CPU engines and $600$\,s on the GPU engines; a case that exceeded the limit is reported as such, and that engine's larger cases in the same sweep were not started. The tables of \Cref{sec:details} instead report medians over repeated calls after one warm-up.

\paragraph{Memory measures.}
All memory readings come from the NVIDIA management library inside the worker. \emph{Occupancy} is the GPU memory in use at the end of each repetition, averaged over the five. It includes the CUDA context and libraries (about $1.4$\,GB, common to both workers) and whatever stays referenced between queries: for \pado\ the compiled program and the sparse matrices cached on the first forward-plus-backward pass for the forward and backward passes (\Cref{sec:evaluation}), for cuPauliProp the per-gate expansions of the last query, which the driver keeps for the backward pass. The \emph{increase over the first pass} is the growth in used memory during the first forward-plus-backward pass, from a baseline holding only the context and, for \pado, the compiled program: the memory a pass needs. The \emph{increase over a later pass} is the same difference over a later pass against a baseline taken just before it, so for \pado\ it excludes the program and its cache, while for cuPauliProp the previous query's expansions are still resident. \pado's memory during compilation, the increase in used memory over \texttt{compile\_program}, is given alongside as an upper bound on the resident program.

\paragraph{Term counts, engines, and precision.}
On the untruncated instances \pado\ and cuPauliProp propagate the same full set of Pauli terms; PauliPropagation.jl runs at the package's default coefficient cutoff of $10^{-10}$ \cite{paulipropagation_jl_repo} and retains fewer at $9$ and $10$ qubits. \pado\ then applies the zero-filtering of \Cref{sec:zero-filter} before evaluation; the surrogate path of PauliPropagation.jl likewise evaluates its set after \texttt{zerofilter!}.

\Cref{tab:bench-engines} gives the engines, versions, and coefficient types: the GPU engines run in single precision and the CPU engines in double, so \pado\ and cuPauliProp are compared at the same precision. cuPauliProp also accepts real single-precision coefficients \cite{cuquantum_cupauliprop}; the harness used its complex single-precision type, twice the size of \pado's \texttt{float32}, and repeated the $w_{\max}=7$ case with the real type (below).

\begin{table}[t]
    \centering
    \small
    \begin{tabular}{l l l}
        \toprule
        Engine & Version & Coefficients \\
        \midrule
        \pado & \texttt{padopauli}~2.0.0, PyTorch~2.11.0 & \texttt{float32} \\
        cuPauliProp & 0.4.0, \texttt{cuquantum-python-cu12}~26.6.0 & \texttt{complex64} \\
        Qiskit pauli-prop & 0.2.1, Qiskit~2.5.2 & \texttt{float64} \\
        PauliPropagation.jl & 0.8.1, Julia~1.12.7 & \texttt{Float64} \\
        \bottomrule
    \end{tabular}
    \caption{The compared engines \cite{cuquantum_cupauliprop,pauli-prop,6vd7-l9bn} and the coefficient type each carries. \pado's \texttt{float32} is a preset override of its \texttt{float64} default; the CPU engines use their package defaults.}
    \label{tab:bench-engines}
\end{table}

\paragraph{Exact reference.}
Every case is also evaluated exactly, without truncation, by a double-precision PennyLane statevector simulation through the reference helpers of \Cref{sec:ex-ref}, which supplies the reference expectation value and, by central differences with $\delta=10^{-4}$, its first four gradient components; the deviations quoted below are absolute differences from these, a gradient deviation being the largest over the four components. The exact statevector baselines of \Cref{sec:bench-stress} use \texttt{pennylane-lightning}~0.45.0 and \texttt{pennylane-lightning-amdgpu}~0.45.0 \cite{asadi2024hybridquantumprogrammingpennylane}.

\paragraph{Numerical agreement.}
On full (untruncated) propagation the engines agree with each other and with the exact statevector value to the resolution the single-precision engines support. On the instances all five engine paths completed ($4$ to $10$ qubits) their values agree to six decimal places, and the two GPU engines agree likewise at $12$ qubits. Against the exact values, the deviation over the five instances is at most $9.6\times10^{-8}$ for \pado, $1.6\times10^{-7}$ for cuPauliProp, and $1.7\times10^{-9}$ for the double-precision CPU engines on the instances they completed. Under an equal max-weight threshold on the $16$-qubit instance, \pado, cuPauliProp, and PauliPropagation.jl return the same value to within $1.3\times10^{-7}$ at every $w_{\max}$ from $2$ to $6$, so \pado's symplectic representation and zero-filtering add no error beyond the truncation itself, which falls from $1.2\times10^{-2}$ at $w_{\max}=2$ to $5.9\times10^{-6}$ at $w_{\max}=6$. Qiskit pauli-prop has no weight threshold and truncates by a term cap instead, so its rows in that sweep are not compared.

\subsection{Evaluation throughput, training, and memory}

\paragraph{Evaluation throughput.}
Against the GPU engine, \pado's evaluation speedup grows with system size, from $2.7\times$ at $4$ qubits to $10.1\times$ at $12$ ($3.5\times$ and $4.3\times$ at $9$ and $10$; \Cref{tab:bench-eval}). One difference is the evaluation working set: at $12$ qubits cuPauliProp processes all $16.8$ million propagated terms on every call, while \pado's evaluation working set holds $4{,}095$ strings. On this untruncated instance both working sets are complete: all $4^{12}-1=16{,}777{,}215$ non-identity strings against all $2^{12}-1=4{,}095$ diagonal ones. The engines also differ in language, kernel design, and data layout, so this comparison does not isolate zero-filtering; \Cref{tab:zf-ablation} does that within \pado.

From $9$ qubits on, the speedups against the CPU engines are larger: at $9$ qubits PauliPropagation.jl and Qiskit pauli-prop are $75\times$ and $160\times$ slower than \pado, at $10$ qubits $165\times$ and $330\times$, and at $12$ qubits none of the three CPU paths completed within the $900$\,s limit. At $4$ qubits the CPU engines are faster (\Cref{tab:bench-eval}); \pado's $5.0$\,ms there, with an evaluation working set of $15$ strings, is a fixed per-query cost, and on the untruncated instances from $6$ qubits \pado\ is the fastest.

On the untruncated scaling instances the surrogate path of PauliPropagation.jl is slower than the library's conventional path at every size it completed (\Cref{tab:bench-eval}), after a one-time graph build and zero-filtering of $1.9$ to $81.3$\,s; on the $16$-qubit sweep it is faster only at $w_{\max}=2$ and $3$ and did not complete $w_{\max}=6$.

\begin{table}[t]
    \centering
    \small
    \begin{tabular}{l r r r r r}
        \toprule
         & \pado & cuPauliProp & Qiskit & PP.jl & PP.jl surrogate \\
        Instance & (GPU) [s] & (GPU) [s] & pauli-prop (CPU) [s] & (CPU) [s] & (CPU) [s] \\
        \midrule
        $4$q  & $0.00497$ & $0.0135$ & $0.00235$ & $0.000553$ & $0.00115$ \\
        $6$q  & $0.00812$ & $0.0229$ & $0.0248$ & $0.0122$ & $0.0225$ \\
        $9$q  & $0.0147$ & $0.0514$ & $2.36$ & $1.11$ & $3.71$ \\
        $10$q & $0.0156$ & $0.0667$ & $5.15$ & $2.57$ & $7.33$ \\
        $12$q & $0.0767$ & $0.772$ & -- & -- & -- \\
        \bottomrule
    \end{tabular}
    \caption{Evaluation time on the untruncated grid Ising instances (protocol of \Cref{sec:bench-setup}); PP.jl (PauliPropagation.jl) on its conventional path and, after a one-time graph build, on its surrogate path. \pado's column excludes its one-time compilation ($0.30$--$8.28$\,s over the rows); -- marks a case that did not complete within the time limit.}
    \label{tab:bench-eval}
\end{table}

\paragraph{Differentiability.}
The backward pass is timed as well. \pado\ and cuPauliProp have a native backward pass (cuPauliProp's through \path{apply_gate_backward_diff} and \path{trace_with_zero_state_backward_diff} \cite{cuquantum_cupauliprop}). PauliPropagation.jl~0.8.1 provides \texttt{rewindgradient}, which returns value and gradient together, so its time is compared with the other engines' forward-plus-backward sum. Qiskit pauli-prop has no backward pass; a parameter-shift workaround would multiply the number of evaluations by twice the parameter count \cite{PhysRevA.99.032331}. For the warm forward-plus-backward pass \pado\ is $3.3\times$ faster than cuPauliProp at $9$ qubits ($0.0467$ against $0.1520$\,s) and $13.3\times$ at $12$ ($0.194$ against $2.576$\,s), and $56\times$ and $133\times$ faster than \texttt{rewindgradient} at $9$ and $10$ qubits ($2.61$ and $6.28$\,s).

\paragraph{Gradient accuracy.}
Against the central difference reference of \Cref{sec:bench-setup}, \pado's gradient deviation lies between $4.8\times10^{-8}$ and $1.1\times10^{-7}$ on every untruncated instance, the resolution of a single-precision engine against a finite-difference reference; \texttt{rewindgradient}'s, in double precision, stays below $7.5\times10^{-8}$. Under an equal max-weight threshold the deviation is the truncation itself: \pado's and cuPauliProp's coincide to three digits and fall from $0.17$ at $w_{\max}=2$ to $1.4\times10^{-4}$ at $w_{\max}=6$. As driven by our harness, cuPauliProp's gradient matches the reference to $1.6\times10^{-7}$ at $4$ and $6$ qubits but deviates by up to $1.8\times10^{-2}$ at $9$ to $12$ qubits and by $3.0$--$3.5\times10^{-2}$ on the instance with batched embedding inputs below, while its expectation values match throughout; we did not trace the cause, and its backward timings are quoted as measured.

\paragraph{Batched embedding inputs.}
\pado\ evaluates a batch of embedding inputs as a single tensor operation; the other engines evaluate it sample by sample. The model is the $3\times3$ grid Ising ansatz at depth $2$, preceded by a layer of per-qubit $X$ rotations whose angles are the embedding inputs ($\mathcal{N}(0,0.3^2)$); it propagates $262{,}003$ terms, of which $511$ remain in \pado's evaluation working set. \pado\ reaches $592$, $2{,}083$, and $5{,}658$ evaluations/s at batch sizes $10$, $100$, and $300$, against $27.5$/s for cuPauliProp ($206\times$ at $300$), which re-propagates in full for every sample, so its rate does not depend on the batch size. With the backward pass included the rates at batch size $300$ are $2{,}060$ training samples/s for \pado\ and $9.3$ for cuPauliProp. The CPU engines, looped the same way, reach forward rates of $0.94$/s and $0.97$/s (Qiskit pauli-prop, batch sizes $10$ and $100$), $1.63$/s (PauliPropagation.jl, batch size $10$), and $0.78$/s at both $10$ and $100$ on the surrogate path, which reuses its graph across samples; the remaining CPU cases (batch size $300$, and PauliPropagation.jl at $100$) exceeded the limit or were not started.

This is the operation a model with embedding inputs repeats throughout training, one batch of inputs per step. The cost is memory: in the forward-plus-backward pass \pado's occupancy grows from $1.9$\,GB at batch size $10$ to $9.2$\,GB at $300$, about $26$\,MB per sample, the intermediates each sample's backward pass reads; a forward pass alone releases them on return. cuPauliProp's occupancy stays at $2.0$\,GB, as it holds one sample at a time.

\begin{table}[t]
    \centering
    \small
    \begin{tabular}{c r r r r r r}
        \toprule
        & \multicolumn{2}{c}{Evaluation [ms]} & \multicolumn{2}{c}{Occupancy [MB]} & & \pado\ compile \\
        \cmidrule(lr){2-3} \cmidrule(lr){4-5}
        $w_{\max}$ & \pado & cuPauliProp & \pado & cuPauliProp & Ratio & [MB] \\
        \midrule
        $2$ & $26.7$ & $87.4$ & $1{,}425$ & $1{,}387$ & $1.0\times$ & $2$ \\
        $3$ & $26.1$ & $89.8$ & $1{,}445$ & $1{,}463$ & $1.0\times$ & $18$ \\
        $4$ & $28.3$ & $103.9$ & $1{,}644$ & $2{,}377$ & $1.4\times$ & $218$ \\
        $5$ & $26.3$ & $152.6$ & $3{,}466$ & $8{,}887$ & $2.6\times$ & $1{,}994$ \\
        $6$ & $43.9$ & $441.9$ & $14{,}810$ & $46{,}494$ & $3.1\times$ & $13{,}108$ \\
        $7$ & $129.9$ & OOM & $74{,}542$ & OOM & -- & $59{,}482$ \\
        \bottomrule
    \end{tabular}
    \caption{The $16$-qubit truncation sweep (protocol of \Cref{sec:bench-setup}). Evaluation, the forward time; Occupancy, the GPU memory in use at the end of a forward-plus-backward pass; Ratio, cuPauliProp's occupancy divided by \pado's; last column, the increase in used memory over \pado's compilation. OOM, out of the $80$\,GB device memory.}
    \label{tab:mw-memory}
\end{table}

\paragraph{Evaluation time and memory under truncation.}
On the $16$-qubit instance, raising $w_{\max}$ from $2$ to $6$ grows the propagated basis from $426$ to $7{,}025{,}029$ terms and multiplies \pado's evaluation time by $1.6\times$ ($26.7$ to $43.9$\,ms) and cuPauliProp's by $5.1\times$ ($87.4$ to $441.9$\,ms), so \pado\ goes from $3.3\times$ to $10.1\times$ faster (\Cref{tab:mw-memory}). PauliPropagation.jl on the CPU takes $3.1$\,ms at $w_{\max}=2$ and $25.9$\,s at $6$, where its coefficient cutoff leaves $3.7$ million of the $7.0$ million terms the GPU engines propagate.

From $w_{\max}=4$ on, \pado\ also occupies less memory in the same sweep, in the forward-plus-backward configuration of \Cref{sec:bench-setup}. Up to $w_{\max}=3$ both occupancies are about $1.4$\,GB, the size of the CUDA context alone; at $w_{\max}=4$, $5$, and $6$ \pado\ occupies $1.6$, $3.4$, and $14.5$\,GB against cuPauliProp's $2.3$, $8.7$, and $45.4$\,GB ($1.4\times$, $2.6\times$, $3.1\times$). cuPauliProp's \texttt{complex64} coefficients ($8$ bytes against \pado's $4$) account for part of the ratio. At $w_{\max}=7$ cuPauliProp runs out of the $80$\,GB device memory with these coefficients and with its real single-precision type alike, while \pado\ completes, occupying $72.8$\,GB and evaluating in $129.9$\,ms. Its gradient there deviates from the exact one by at most $1.6\times10^{-5}$, against $1.4\times10^{-4}$ at $w_{\max}=6$, while the expectation value had already converged ($6.3\times10^{-6}$ at $w_{\max}=7$ and $5.9\times10^{-6}$ at $w_{\max}=6$).

The two engines hold different data. At $w_{\max}=6$ \pado's occupancy above the context is mainly the resident compiled program, $6.4$\,GB, and the forward and backward matrices its first forward-plus-backward pass builds and keeps, $6.5$\,GB of a $6.6$\,GB increase over that pass. The cuPauliProp driver holds the per-gate expansions of the last query for the backward pass. A first forward-plus-backward pass raises the used memory by $6.6$\,GB for \pado\ and $43.4$\,GB for cuPauliProp ($6.6\times$); a later pass by $234$\,MB and $9{,}409$\,MB, the latter with $34.9$--$35.4$\,GB of the previous query's expansions still resident.

Apart from $w_{\max}=7$, the untruncated $12$-qubit instance is the one case that fills the device. There the first forward-plus-backward pass raises the used memory by $78.2$\,GB for cuPauliProp against $40.5$\,GB for \pado\ ($1.9\times$; $39.9$\,GB of \pado's increase stays resident beside its $23.7$\,GB program), and later passes add $2.3$\,GB (cuPauliProp) and $1.7$\,GB (\pado). With occupancies of $79.5$\,GB and $66.4$\,GB on an $80$\,GB device, both engines are near capacity.

\paragraph{Compilation cost against cuPauliProp.}
On this workload and platform the trade-off is \pado's up-front compilation, $8.28$\,s and $36.6$\,GB of memory during compilation at $12$ qubits, which makes cuPauliProp preferable for a single evaluation and \pado\ preferable whenever a circuit is evaluated many times, as in training. At $12$ qubits the compilation is amortized after about $12$ forward evaluations or $3.5$ forward-plus-backward passes relative to cuPauliProp's per-query cost, or about $14$ passes once \pado's first forward-plus-backward pass ($24.4$\,s) is counted as well.

\subsection{Stress test vs.\ exact statevector}
\label{sec:bench-stress}

The grid Ising workload is an easy case for Pauli propagation: on its $16$-qubit instance the truncation error is $5.9\times10^{-6}$ at $w_{\max}=6$. For a harder regime, and a comparison against an exact statevector simulator, we turn to a variant of the random circuit family of \cite{Hirviniemi:2026dzs} with fixed density and fixed depth. The circuit is a randomly drawn block $U(\bm\phi)$ followed by a trainable brickwork ansatz $V(\bm\theta)$. $U$ applies three layers of a per-qubit $R_X(\phi)$ rotation followed by a $\mathrm{CZ}$ layer on an independent Erd\H{o}s--R\'enyi graph (edge probability $0.15$), then a per-qubit $R_X(\phi)$ and a per-qubit $R_Y(\phi)$ layer; $V$ applies two layers of a per-qubit $R_Y(\theta),R_Z(\theta)$ pair followed by a brickwork $\mathrm{CZ}$ pattern with alternating offset.

The embedding angles are $\phi\sim\mathcal N(0,\tau^2)$ with $\tau^2=0.24$ and the trainable angles are initialized from $\mathcal N(0,0.1^2)$; $\bm\phi$ enters \pado\ through \texttt{embedding} and $\bm\theta$ through \texttt{thetas}, and we read out $\langle Z_0\rangle$ and $\langle Z_{n-2}Z_{n-1}\rangle$. We sweep $n=12,16,20$ at fixed $w_{\max}=9$ under the \texttt{gpu} preset in \texttt{float32}, against exact PennyLane baselines: \texttt{lightning.qubit} on the host CPU, the accuracy reference up to $n=16$, and \texttt{lightning.amdgpu} on the same MI300X, the accuracy reference at $n=20$ and the timed baseline at every size.

\paragraph{Accuracy on the random circuit.}
On this random circuit the truncation error against the exact value, over a batch of $100$ random embedding inputs, has a small mean but a worst case more than ten times the mean (\Cref{tab:stress-acc}). At $w_{\max}=9$ the mean absolute error ($0.0008$, $0.0031$, $0.0180$) and the worst case over the batch ($0.009$, $0.044$, $0.204$) both grow with system size ($n=12$, $16$, $20$); both are absolute deviations of $\langle Z_0\rangle$ and $\langle Z_{n-2}Z_{n-1}\rangle$, which lie in $[-1,1]$. Raising the threshold lowers the error at a cost: at $n=16$ the step $w_{\max}=9\!\to\!10$ more than halves the worst-case error, from $0.044$ to $0.019$, while the propagated basis grows from $0.72$ to $2.60$ million terms. On this instance at $n=20$ the approximation is no longer a reliable estimator of the exact value.

\begin{table}[t]
    \centering
    \small
    \begin{tabular}{c c r r r r}
        \toprule
         & & & & \multicolumn{2}{c}{absolute error} \\
        \cmidrule(lr){5-6}
        $n$ & $w_{\max}$ & Propagated terms & Zero-filtered terms & mean & max \\
        \midrule
        $12$ & $9$ & $0.25$M  & $853$      & $0.0008$ & $0.009$ \\
        $16$ & $9$ & $0.72$M  & $2{,}088$  & $0.0031$ & $0.044$ \\
        $16$ & $10$ & $2.60$M  & $3{,}867$  & $0.0016$ & $0.019$ \\
        $20$ & $9$ & $0.80$M  & $2{,}485$  & $0.0180$ & $0.204$ \\
        \bottomrule
    \end{tabular}
    \caption{Truncation accuracy on the random circuit of this stress test against an exact statevector baseline (\texttt{lightning.qubit} up to $n=16$, \texttt{lightning.amdgpu} at $n=20$) over $100$ random embedding inputs; M denotes $10^{6}$.}
    \label{tab:stress-acc}
\end{table}

\paragraph{Forward throughput vs.\ statevector.}
The timed operation is the forward pass of a model with embedding inputs: one circuit evaluated across a batch of them, which the statevector baseline simulates one sample at a time. \Cref{tab:stress-tp} reports forward throughput after the one-time compilation, for batches of $128$--$512$ embedding inputs. Compilation takes $0.45$ and $0.61$\,s at $n=16$ and $20$. \pado\ reaches thousands to tens of thousands of evaluations per second and speeds up further with batch size, consistent with a fixed per-query cost amortized over the batch, whereas the baseline's rate is nearly independent of the batch size. The speedup therefore grows with batch size at every $n$, at $n=16$ from $57\times$ to $194\times$ over \texttt{lightning.amdgpu} as the batch grows from $128$ to $512$.

The same ordering holds with the backward pass included. After the one-time compilation, an Adam loop of $10$ steps that minimizes the batch mean of $\langle Z_0\rangle$ over the trainable angles (batch of $16$ embedding inputs, learning rate $0.05$) runs $9.5$--$15.2\times$ faster than the same loop run one sample at a time with \texttt{lightning.amdgpu}'s \texttt{adjoint} differentiation over $n=12$--$20$.

The compiled program's loss and the exact circuit's loss, each along its own trajectory, agree to within $0.004$ at every step ($0.552\to0.238$ against $0.553\to0.242$ at $n=20$). On this instance the approximation is an imperfect per-sample estimator whose training loss nonetheless tracks the exact circuit's.

\begin{table}[t]
    \centering
    \small
    \begin{tabular}{c c r r}
        \toprule
        $n$ & Batch & \pado\ (eval/s) & \texttt{lightning.amdgpu} (eval/s) \\
        \midrule
        \multirow{3}{*}{$12$} & $128$ & $11{,}679$ & $197.1$ ($59\times$) \\
                              & $256$ & $22{,}645$ & $197.2$ ($115\times$) \\
                              & $512$ & $36{,}867$ & $201.1$ ($183\times$) \\
        \midrule
        \multirow{3}{*}{$16$} & $128$ & $7{,}451$  & $130.8$ ($57\times$) \\
                              & $256$ & $15{,}315$ & $130.8$ ($117\times$) \\
                              & $512$ & $25{,}229$ & $130.1$ ($194\times$) \\
        \midrule
        \multirow{3}{*}{$20$} & $128$ & $5{,}223$  & $59.6$ ($88\times$) \\
                              & $256$ & $10{,}638$ & $57.7$ ($184\times$) \\
                              & $512$ & $17{,}747$ & $57.8$ ($307\times$) \\
        \bottomrule
    \end{tabular}
    \caption{Throughput of the forward pass (evaluations per second) on the stress test circuit after a one-time compilation: \pado\ (one batched tensor operation; $w_{\max}=9$, \texttt{gpu} preset, \texttt{float32}, MI300X; median of five calls after one warm-up) against the statevector baseline \texttt{lightning.amdgpu} (\texttt{complex128}) on the same GPU, called once per sample in one timed pass over the batch after one warm-up sample; parentheses give the speedup.}
    \label{tab:stress-tp}
\end{table}

On the grid Ising workload \pado\ evaluates faster than the GPU engine on every scaling instance (\Cref{tab:bench-eval}) and occupies less GPU memory under truncation at $w_{\max}\ge4$ (\Cref{tab:mw-memory}). On the random circuit stress test, after a one-time compilation, the batched forward pass runs $57$--$307\times$ faster than the per-sample GPU statevector loop used here over the sizes and batch sizes measured (\Cref{tab:stress-tp}), and a training loop of ten steps $9.5$--$15.2\times$ faster than the same loop run one sample at a time with GPU \texttt{adjoint} differentiation. There, however, accuracy is the limiting factor: at $w_{\max}=9$ the worst error over the data batch reaches $0.204$ at $n=20$ (\Cref{tab:stress-acc}). In both comparisons the gain is measured after the one-time compilation, so it applies when the same circuit is queried many times, as in training.

\section{Applications}
\label{sec:examples}

Each case below uses the interface of \Cref{sec:compile-once,sec:api}. The listings omit setup and plotting code.

\subsection{Example at utility scale: 127-qubit kicked Ising}
\label{sec:ex-kicked-ising}

We simulate the $127$-qubit kicked transverse-field Ising experiment on IBM's Eagle heavy-hex lattice, the basis of the claim of quantum utility of \cite{Kim2023Evidence} and since then a common target for classical Pauli propagation methods \cite{doi:10.1126/sciadv.adk4321,PhysRevLett.133.120603,Rudolph2023ClassicalSS,6vd7-l9bn,pauli-prop}. Each Trotter step applies a transverse-field layer $\prod_{q} e^{-i h X_q/2}$ followed by an Ising layer $\prod_{(u,v)\in E} e^{\,i(\pi/4) Z_uZ_v}$ (that is, $R_{ZZ}(-\pi/2)$, which is Clifford) on the $144$ heavy-hex edges, so the $X$ rotations are the only non-Clifford gates. \pado\ represents the $R_{ZZ}(-\pi/2)$ gates as Pauli rotations, whose cosine branch carries the coefficient $\cos(-\pi/2)=0$ and is removed by the coefficient truncation at the build point, so only the $X$ rotations add terms. We propagate the single-site magnetization $\langle Z_{62}\rangle$ through $20$ Trotter steps and sweep the field $h\in[0,\pi/2]$.

The sweep uses the observable and truncation settings of the utility example distributed with PauliPropagation.jl \cite{paulipropagation_jl_repo}: $w_{\max}=8$ and a coefficient threshold of $10^{-4}$. \Cref{lst:kicked} builds the circuit and sweeps the field. The coefficient threshold is applied at a build point that moves with $h$, so the listing compiles once per field value.

A $9$-qubit patch of the lattice, run untruncated, matches an exact PennyLane statevector to $2.2\times10^{-14}$, so the gate conventions are right.

\begin{figure}[t]
    \centering
    \includegraphics[width=\linewidth]{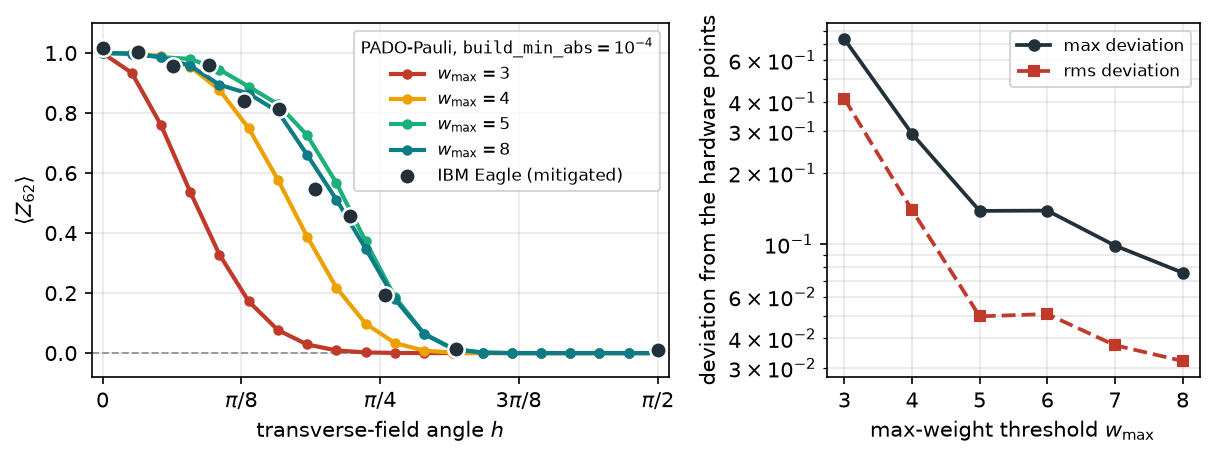}
    \caption{Single-site magnetization $\langle Z_{62}\rangle$ of the $127$-qubit kicked Ising on the IBM Eagle heavy-hex lattice against the transverse-field angle $h$. \emph{Left:} four of the six swept max-weight thresholds against the error-mitigated hardware points of \cite{Kim2023Evidence} ($w_{\max}=6$ and $7$ omitted). \emph{Right:} maximum and root-mean-square deviation from those eleven points at every swept threshold.}
    \label{fig:kicked-ising}
\end{figure}

No exact statevector is feasible at this size, so we apply the apparent convergence check of \Cref{sec:trunc-accuracy}: we repeat the whole sweep at every $w_{\max}$ from $3$ to $8$ with the coefficient threshold held at $10^{-4}$. \Cref{fig:kicked-ising} overlays four of the six settings on the error-mitigated hardware points of \cite{Kim2023Evidence}, as distributed with \cite{paulipropagation_jl_repo}; the deviations below are taken from the curve linearly interpolated from the twenty field values of the sweep to the eleven hardware angles.

At $w_{\max}\ge5$ the magnetization stays near $1$ at small field, falls through a sharp crossover near $h=\pi/4$, and vanishes by $h\approx 3\pi/8$. At $h=0$ every $X$ rotation is the identity and $\langle Z_{62}\rangle=1$ exactly; at $h=\pi/2$ every gate is Clifford, the surviving evaluation working set is empty, and the program returns $0$.

Raising $w_{\max}$ from $3$ to $5$ shifts the crossover to larger $h$, and $w_{\max}=7$ and $8$ move it slightly back; the deviation falls overall, though not monotonically: the maximum absolute deviation runs $0.738$, $0.293$, $0.138$, $0.139$, $0.099$, $0.076$ over $w_{\max}=3$ to $8$, and the root-mean-square deviation $0.413$, $0.140$, $0.050$, $0.051$, $0.037$, $0.032$.

Between $w_{\max}=5$ and $6$ the curve changes by at most $0.010$, so the apparent convergence check would stop there; yet the next step, $6$ to $7$, moves the curve by $0.044$. Apparent convergence alone is therefore not a sufficient stopping criterion.

One run of the whole sweep (compilation plus one evaluation at each of the twenty fields) takes $105$ to $215$\,s of wall time and peaks at $0.08$ to $0.63$\,GB of GPU memory over the six settings of $w_{\max}$, on a single AMD Instinct MI300X.

\begin{lstlisting}[caption={127-qubit kicked Ising magnetization sweep on the IBM Eagle heavy-hex lattice.},label={lst:kicked}]
n, n_steps = 127, 20
edges = eagle_edges()                   # 144 heavy-hex edges (0-indexed)

qc = Circuit(n)                         # RX(h): param 0 ; RZZ(-pi/2): param 1
for _ in range(n_steps):
    for q in range(n): qc.rx(q, param_idx=0)
    for (u, v) in edges: qc.rzz(u, v, param_idx=1)

evs = []
for h in np.linspace(0, np.pi / 2, 20):                      # transverse-field sweep
    thetas = torch.tensor([h, -np.pi / 2], dtype=torch.float64)
    program = qc.compile(
        observables=[("Z", [62])], preset="gpu",             # <Z_62>
        dtype="float32", max_weight=8,
        build_thetas=thetas, build_min_abs=1e-4)
    evs.append(float(program.expvals(thetas)[0]))
\end{lstlisting}

\subsection{Variational quantum eigensolver (VQE)}

This chemistry case, a variational quantum eigensolver \cite{Peruzzo2014}, adds two elements: an observable with many terms assembled from an external Hamiltonian, and a training loop. The system is $\mathrm{H}_2$ in the STO-3G basis, with the qubit Hamiltonian from \texttt{qml.qchem.molecular\_hamiltonian} under a Jordan--Wigner mapping: $4$ qubits and $15$ Pauli terms at the experimental equilibrium bond length $R=0.7414$\,\AA. The identity term (coefficient $-0.098864$\,Ha) is held out as a classical offset, added back after evaluation, leaving $14$ terms in the \texttt{PauliSum}.

The ansatz is minimal: two $X$ gates prepare the Hartree--Fock determinant $\ket{1100}$ and one rotation on four qubits, $\exp(-i\theta\, X_0X_1X_2Y_3/2)$, supplies the double excitation, giving $3$ gates and $1$ trainable angle. Compilation is untruncated in \texttt{float64}; the propagated basis is $18$ terms, of which zero-filtering retains $12$. \Cref{lst:vqe} assembles the observable, builds the ansatz, and compiles the program.

\begin{lstlisting}[caption={$\mathrm{H}_2$ VQE: observable with many terms, minimal ansatz, gradient training; the omitted training loop follows \Cref{lst:hybrid}.},label={lst:vqe}]
from padopauli import Circuit, PauliSum

# H (15 Pauli terms) from PennyLane; the identity term becomes a classical offset.
obs = PauliSum(4)
for c, word, qubits in non_identity_terms:      # 14 terms
    obs.add_from_str(word, float(c), qubits=qubits)

qc = Circuit(4)
qc.x(0); qc.x(1)                                # Hartree-Fock |1100>
qc.rot("XXXY", [0, 1, 2, 3], param_idx=0)       # double excitation
program = qc.compile(observables=[obs], preset="gpu", dtype="float64")

theta = torch.zeros(1, dtype=torch.float64, requires_grad=True)
# then 300 Adam steps on energy = program.expvals(theta)[0] + e_offset
\end{lstlisting}

\begin{figure}[t]
    \centering
    \includegraphics[width=0.68\linewidth]{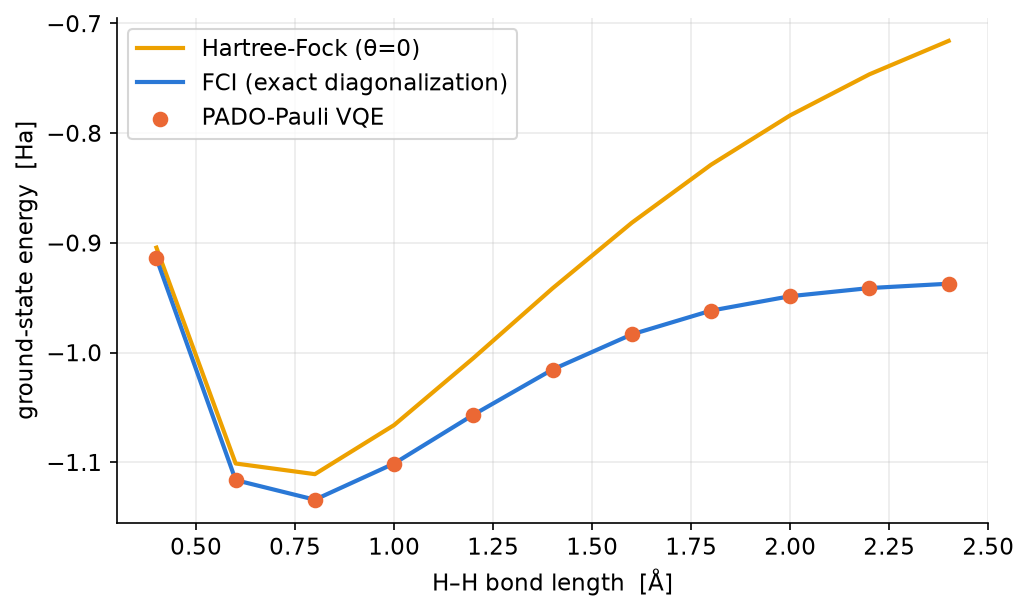}
    \caption{$\mathrm{H}_2$ dissociation curve in the STO-3G basis: a program compiled and trained at each of the eleven bond lengths, against the FCI value from exact diagonalization and the Hartree--Fock reference ($\theta=0$).}
    \label{fig:vqe-h2}
\end{figure}

Starting from $\theta=0$ (the Hartree--Fock energy, $-1.11668439$\,Ha), $300$ Adam steps at learning rate $0.05$ converge to $-1.1372701749$\,Ha at $\theta^\ast=-0.226136$. This differs from the exact ground energy of the $16\times16$ Hamiltonian matrix (the full configuration interaction value, FCI in \Cref{fig:vqe-h2}) by $-2.22\times10^{-16}$\,Ha, the spacing between adjacent double-precision numbers at this energy. Repeating the procedure at eleven bond lengths from $0.4$ to $2.4$\,\AA\ ($250$ steps each) reproduces the dissociation curve with a maximum deviation of $1.7\times10^{-12}$\,Ha (\Cref{fig:vqe-h2}).

The agreement has a structural reason: the exact ground state of $\mathrm{H}_2$ in this basis lies in the span of $\ket{1100}$ and $\ket{0011}$, the one-parameter ansatz reaches it, and no truncation is applied, so the compiled program represents the circuit exactly. The case validates observable assembly, compilation, evaluation, and gradient end to end against an independently known value.

\subsection{MaxCut-\qaoa}
\label{sec:ex-maxcut}

MaxCut is a natural benchmark because its cost Hamiltonian is a sparse sum of two-local $Z_iZ_j$ terms. Maximizing the cut on a graph $G=(V,E)$ is equivalent to minimizing $\sum_{(i,j)\in E}\langle Z_iZ_j\rangle$, and $\mathbb{E}[\mathrm{cut}]=(|E|-\sum\langle Z_iZ_j\rangle)/2$. A depth-$p$ \qaoa\ ansatz \cite{farhi2014quantumapproximateoptimizationalgorithm} alternates a cost block of $Z_iZ_j$ rotations with a mixer block of $X$ rotations after a Hadamard layer that prepares $\ket{+}^{\otimes n}$.

The instance is a deterministic $15$-vertex graph with $30$ edges (a ring plus a chord from every vertex to the one seven positions away) at depth $p=4$ with one shared angle per block, so the $180$ rotations among its $195$ gates share $2p=8$ trainable angles. The angles start from a linear ramp with $\delta t=0.8$, so that $\gamma_i=(i/p)\,\delta t$ and $\beta_i=-(1-i/p)\,\delta t$, and are trained for $200$ Adam steps at learning rate $0.05$ under the \texttt{gpu} preset at $w_{\max}=5$. Zero-filtering reduces the program from $335{,}602$ propagated terms to the $1{,}923$ diagonal terms that contribute ($0.57\%$). \Cref{lst:qaoa} assembles the objective, trains it, and samples the result.

\begin{lstlisting}[caption={MaxCut-\qaoa\ objective, training, and sampling; the omitted training loop follows \Cref{lst:hybrid}.},label={lst:qaoa}]
from padopauli import Circuit, pennylane_sample

p, dt = 4, 0.8
qc = Circuit(n_qubits)
for q in range(n_qubits): qc.h(q)                            # |+>^n
for i in range(p):
    for (u, v) in edges: qc.rzz(u, v, param_idx=2 * i)       # cost: one shared gamma
    for q in range(n_qubits): qc.rx(q, param_idx=2 * i + 1)  # mixer: one shared beta

# The whole edge objective is ONE observable with many terms: the list of term tuples
# is passed as a single entry of observables.
cost_terms = [("ZZ", [u, v], 1.0) for (u, v) in edges]
qc.compile(observables=[cost_terms], preset="gpu", max_weight=5)

# Linear ramp: gamma_i = (i/p) dt, beta_i = -(1 - i/p) dt
theta0 = [a for i in range(1, p + 1) for a in ((i / p) * dt, -(1 - i / p) * dt)]
thetas = torch.nn.Parameter(torch.tensor(theta0, dtype=torch.float64))
# then 200 Adam steps minimizing qc.expvals(thetas)[0], the whole cost as one observable;
# best_thetas holds the trained angles.

samples = pennylane_sample(circuit=qc.gates, thetas=best_thetas.cpu(),
                           n_qubits=n_qubits, shots=4000, seed=0)
\end{lstlisting}

\begin{figure}[t]
    \centering
    \includegraphics[width=0.7\linewidth]{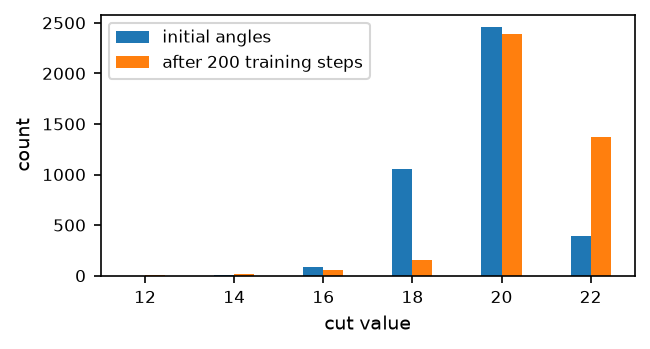}
    \caption{Histograms of the cut values of $4{,}000$ bitstrings drawn from the MaxCut-\qaoa\ circuit of \Cref{lst:qaoa} at the initial angles (blue) and after $200$ training steps (orange, at the trained angles), both sampled from the exact statevector baseline.}
    \label{fig:qaoa-hist}
\end{figure}

Training drives the compiled objective from $\sum_{(i,j)\in E}\langle Z_iZ_j\rangle=-8.93$ at the initial angles to $-11.15$, an expected cut of $20.57$ against the optimum of $22$ found by enumerating the $2^{15}$ bitstrings. To measure the effect on the circuit itself, \path{pennylane_sample} draws $4{,}000$ shots from the circuit at the initial angles and another $4{,}000$ at the trained angles. In \Cref{fig:qaoa-hist} the initial circuit samples a mean cut of $19.57$ and $9.8\%$ of its shots reach the optimum; the trained circuit samples a mean cut of $20.52$, and $34.4\%$ of its shots reach the optimum. For reference, exactly $30$ of the $32{,}768$ bitstrings are optimal, so uniform sampling hits one with probability $0.092\%$ and has a mean cut of $|E|/2 = 15$.

The compiled objective is accurate at both points: its $\sum\langle Z_iZ_j\rangle$ differs from the exact statevector value by $0.13$ at the initial angles and $0.036$ at the trained angles.

\subsection{SAFE ma-\qaoa}
\label{sec:ex-safe}

The final case uses the compiled program as the \emph{surrogate} of the surrogate-assisted and fine-tuning enhanced (SAFE) method \cite{kim2026safemaqaoasurrogateassistedfinetuning}, a cheap source of starting parameters for an expensive optimizer. For the multi-angle \qaoa\ (ma-\qaoa) ansatz \cite{Herrman2022}, whose parameterization with one angle per gate is expressive at low depth but expensive to train \cite{kim2026safemaqaoasurrogateassistedfinetuning}, SAFE runs three stages: a low-weight Pauli propagation (LWPP) warm-up on a truncated \pado\ program, a parameter distillation step, and a fine-tune on the exact objective. SAFE aims to reduce the number of exact optimizer steps.

The instance is an Erd\H{o}s--R\'enyi MaxCut graph on $n=16$ vertices at edge probability $0.3$ with $33$ unweighted edges; the objective is to minimize $H_C=\sum_{(i,j)\in E}Z_iZ_j$. Enumerating the $2^{16}$ configurations gives $C_{\min}=-19$ and $C_{\max}=+33$ for $C=\langle H_C\rangle$, and quality is reported as the approximation ratio $\alpha(C)=(C_{\max}-C)/(C_{\max}-C_{\min})$. The ansatz is ma-\qaoa\ at depth $p=2$: a Hadamard layer, then per layer one $R_{ZZ}$ per edge and one $R_X$ per qubit, each with its own \texttt{param\_idx}, for $114$ gates with $98$ trainable angles, compared with the $2p$ shared angles of \Cref{sec:ex-maxcut}. We compare SAFE with a cold start; both begin at one of the constant initializations of \cite{kim2026safemaqaoasurrogateassistedfinetuning}, every cost angle at $+0.01$ and every mixer angle at $-0.01$, an initialization with $\alpha=0.6350$. \Cref{lst:safe} shows the warm-up stage.

\begin{lstlisting}[float=tb,caption={SAFE ma-\qaoa: a truncated \pado\ warm-up supplying the starting point for an exact fine-tune; the omitted training loops follow \Cref{lst:hybrid}.},label={lst:safe}]
qc = build_ma_qaoa(n, edges, p)         # one param_idx per gate
obs = [("ZZ", [u, v], 1.0) for (u, v) in edges]

# stage 1: the surrogate, truncated at max weight 4, differentiable, on the GPU
lwpp = qc.compile(observables=[obs], preset="gpu", dtype="float64", max_weight=4)
theta = torch.nn.Parameter(theta_init.clone())   # (1, 98): the kept batch axis
# then 500 Adam steps on lwpp.expvals(theta)[:, 0].sum()

# stage 3: the same angles, now refined on the exact objective
exact = make_exact(qc.gates)            # PennyLane default.qubit, torch backprop
\end{lstlisting}

The surrogate's compiled program is small: $335$ of the $18{,}136$ propagated terms survive zero-filtering, and its $114$ steps hold $18{,}264$ nonzero entries in total. Evaluated on the same gate list, \pado's PennyLane reference helpers and the \texttt{make\_exact} helper of \Cref{lst:safe} agree to $3.2\times10^{-16}$, so \texttt{make\_exact} uses the engine's gate and angle conventions.

Stage one runs $500$ Adam steps at learning rate $0.02$ on the surrogate alone, with no exact evaluation during optimization, and reaches $\alpha=0.9550$ evaluated exactly at the angles trained on the surrogate. Stage two deletes every rotation whose trained angle satisfies $|\theta|\le0.3$, removing $47$ of the $98$ rotations and leaving $67$ gates on $51$ parameters. Stage three fine-tunes on the exact objective at learning rate $0.02$; the surrogate is not used again.

\begin{figure}[t]
    \centering
    \includegraphics[width=0.68\linewidth]{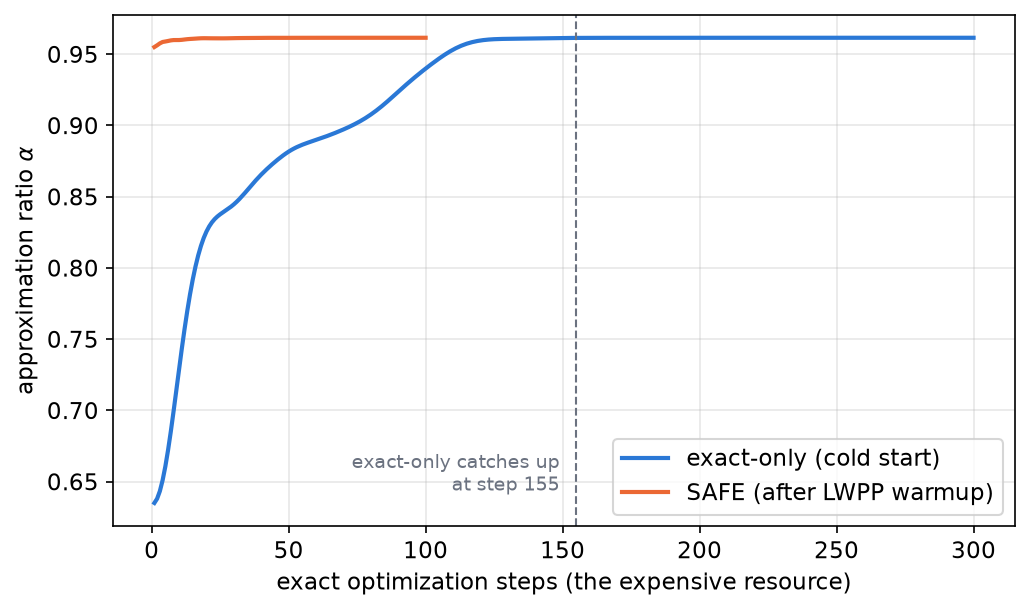}
    \caption{Approximation ratio vs.\ the number of exact optimizer steps on the $n=16$ MaxCut instance at $p=2$: cold start (blue) against the SAFE run after the LWPP warm-up (orange, its $100$ exact steps without distillation, on all $98$ angles). The dashed line at step $155$ marks where the cold start first comes within $10^{-4}$ of SAFE's value after $100$ steps.}
    \label{fig:safe-convergence}
\end{figure}

After $100$ exact steps SAFE reaches $\alpha=0.9615$ with or without distillation. The cold start reaches $0.9399$ in $100$ steps and needs $155$ to match SAFE's value at $100$ (\Cref{fig:safe-convergence}). SAFE reaches that value in $1.6\times$ fewer exact optimizer steps ($100$ against $155$) and, with distillation, on roughly half the parameters. The SAFE paper \cite{kim2026safemaqaoasurrogateassistedfinetuning} evaluates the reduction in exact steps across problem families and depths.

\section{Conclusion}
\label{sec:conclusion}

We presented \pado, a GPU-accelerated and differentiable implementation of the Pauli propagation surrogate. Its organizing idea is a split between compile time and query time: everything that does not change between queries (propagation, max-weight truncation, zero-filtering, and coefficient truncation at a fixed build point) is done once, and everything that depends on the queried parameters is a fixed pass over sparse matrices. Two further properties hold. First, because the compiled program is a sparse tensor program whose backward pass (\texttt{vjp}, the default) is registered with PyTorch's automatic differentiation, it composes with surrounding differentiable layers. Second, because zero-filtering removes only branches that cannot contribute, the filtered program returns the same expectation values; in the ablation, zero-filtering reduces steady-state memory by $1.4$--$1.5\times$ from $w_{\max}=5$ on and speeds up the batched forward pass by about $1.5\times$.

Apart from floating-point round-off, truncation is the only approximation and is an explicit control; the compiled program is an approximate simulator whose error must be checked per instance (\Cref{sec:trunc-cost}, \Cref{sec:bench-stress}). Its intended regime is repeated evaluation with gradients and batched queries, as in the training loops of \Cref{sec:examples}.

\paragraph{Limitations.}
The approach is not a general-purpose simulator. \emph{(i) Truncation is not universally safe.} Pauli propagation suits circuits and observables whose Pauli terms stay at low weight, or circuits with genuine noise; for deep circuits with many non-Clifford rotations at large angles, which spread the observable onto many Pauli strings of high weight, the discarded coefficient mass $\Delta$ need not be small, and sufficient conditions for reliable simulation, with instances where the method fails, have been given in terms of the noise rate and the fraction of T gates in circuits of Clifford and T gates \cite{GonzalezGarcia2025paulipath}. \emph{(ii) The provable guarantees are conditional.} The tightest bounds assume Pauli noise \cite{Fontana2025Classical,PhysRevLett.133.120603} or randomly drawn, locally scrambling layers \cite{lh6x-7rc3}; for a specific noiseless structured instance only the worst-case $\ell_1$ bound applies, and it can be loose. \emph{(iii) Compilation is a fixed up-front cost,} so for a program queried only a few times the compilation cost is not amortized, and the first forward-plus-backward pass can raise device memory use by more than compilation needed ($40.5$ against $36.6$\,GB on the untruncated $12$-qubit instance of \Cref{sec:benchmarks}). Coefficient truncation also ties the program to its build point, so a parameter sweep recompiles (\Cref{sec:ex-kicked-ising}). \emph{(iv) Highly branching gates.} Unitaries with many controls make the Pauli representation prohibitively large \cite{6vd7-l9bn}. \emph{(v) Accuracy at scale is only partially characterized.} Every accuracy claim here is measured against an exact baseline at $n\le28$; beyond that the evidence is indirect, and the $127$-qubit sweep's deviation from the hardware points was still falling at the largest $w_{\max}$ measured.

\paragraph{Outlook.}
Two extensions follow. The accuracy study of \Cref{sec:trunc-cost} should be extended to sizes where exact simulation is infeasible, with the discarded mass $\Delta$ of coefficient truncation reported as an error certificate. Moving step matrices to host memory automatically when a compilation fills the device would extend the reachable qubit count without the evaluation cost that the \texttt{hybrid} preset adds to every query.

\section*{Code and Data Availability}
\pado~2.0.0, the version used for every result in this paper, is distributed as the Python package \texttt{padopauli} through PyPI and \url{https://github.com/Norma-Q/PADO-Pauli}. It installs with the following commands:
\begin{lstlisting}[language=bash]
pip install padopauli                # the engine
pip install "padopauli[reference]"   # adds PennyLane for the exact cross-checks
\end{lstlisting}
The supported Python and PyTorch versions are listed in the repository. The package is licensed under the NORMA Binary License~1.0, which permits noncommercial use free of charge; commercial use requires a separate agreement. The notebooks, the measurement scripts, and the recorded data behind every table and figure are in the same repository under the Apache License~2.0; the snapshot used for this paper (tag \texttt{v2.0.0-whitepaper}) is archived at \url{https://doi.org/10.5281/zenodo.23310965}.

The comparisons with cuPauliProp, Qiskit pauli-prop, and PauliPropagation.jl in \Cref{sec:benchmarks} ran on an NVIDIA A100, since cuPauliProp requires CUDA; every other result ran on an AMD Instinct MI300X (\Cref{tab:platforms}).

\begin{table}[h]
    \centering
    \small
    \begin{tabular}{l l >{\raggedright\arraybackslash}p{5cm}}
        \toprule
         & AMD Instinct MI300X & NVIDIA A100 80GB PCIe \\
        \midrule
        Results & all other results & comparisons with the three Pauli propagation engines \\
        Device memory & $191.7$\,GB & $80$\,GB \\
        Host CPU & Intel Xeon Platinum 8568Y+, $20$ vCPUs & Intel Xeon Gold 6338, $24$ vCPUs \\
        PyTorch & 2.11.0 (ROCm 7.2.4) & 2.11.0 (CUDA 12.6) \\
        Python & 3.11.15 & 3.11.15 \\
        Exact reference & PennyLane 0.45.1 & PennyLane 0.45.1 \\
        \bottomrule
    \end{tabular}
    \caption{Measurement platforms.}
    \label{tab:platforms}
\end{table}

\section*{Acknowledgments}

The authors gratefully acknowledge NORMA, Inc. for its generous support of this research. The authors used Claude (Anthropic) for writing code and for editing the manuscript. The authors verified all of Claude's output.

{\small
\bibliographystyle{unsrt}
\bibliography{main}

\begin{thebibliography}{10}

\bibitem{Gottesman1998TheHR}
Daniel Gottesman.
\newblock The {Heisenberg} representation of quantum computers.
\newblock In S.~P. Corney, R.~Delbourgo, and P.~D. Jarvis, editors, {\em
  Group22: Proceedings of the XXII International Colloquium on Group
  Theoretical Methods in Physics}, pages 32--43, Cambridge, MA, 1999.
  International Press.
\newblock arXiv:quant-ph/9807006.

\bibitem{PhysRevA.99.062337}
Patrick Rall, Daniel Liang, Jeremy Cook, and William Kretschmer.
\newblock Simulation of qubit quantum circuits via {Pauli} propagation.
\newblock {\em Phys. Rev. A}, 99:062337, Jun 2019.

\bibitem{6vd7-l9bn}
Manuel~S. Rudolph, Tyson Jones, Yanting Teng, Armando Angrisani, and Zo\"e
  Holmes.
\newblock Pauli propagation: A computational framework for simulating quantum
  systems.
\newblock {\em PRX Quantum}, 7:032001, Aug 2026.

\bibitem{begusic2023fastsim}
Tomislav Begu{\v{s}}i{\'c} and Garnet Kin-Lic Chan.
\newblock Fast classical simulation of evidence for the utility of quantum
  computing before fault tolerance, 2023.
\newblock arXiv:2306.16372 [quant-ph].

\bibitem{doi:10.1126/sciadv.adk4321}
Tomislav Begušić, Johnnie Gray, and Garnet Kin-Lic Chan.
\newblock Fast and converged classical simulations of evidence for the utility
  of quantum computing before fault tolerance.
\newblock {\em Science Advances}, 10(3):eadk4321, 2024.

\bibitem{PhysRevLett.133.120603}
Yuguo Shao, Fuchuan Wei, Song Cheng, and Zhengwei Liu.
\newblock Simulating noisy variational quantum algorithms: A polynomial
  approach.
\newblock {\em Phys. Rev. Lett.}, 133:120603, Sep 2024.

\bibitem{Rudolph2023ClassicalSS}
Manuel~S. Rudolph, Enrico Fontana, Zo\"e Holmes, and Lukasz Cincio.
\newblock Classical surrogate simulation of quantum systems with {LOWESA},
  2023.
\newblock arXiv:2308.09109 [quant-ph].

\bibitem{PRXQuantum.5.010308}
Joseph Tindall, Matthew Fishman, E.~Miles Stoudenmire, and Dries Sels.
\newblock Efficient tensor network simulation of {IBM}'s {Eagle} kicked {Ising}
  experiment.
\newblock {\em PRX Quantum}, 5:010308, Jan 2024.

\bibitem{liao2023simulationibmskickedising}
Hai-Jun Liao, Kang Wang, Zong-Sheng Zhou, Pan Zhang, and Tao Xiang.
\newblock Simulation of {IBM}'s kicked {Ising} experiment with projected
  entangled pair operator, 2023.
\newblock arXiv:2308.03082 [quant-ph].

\bibitem{PhysRevResearch.6.013326}
Siddhartha Patra, Saeed~S. Jahromi, Sukhbinder Singh, and Rom\'an Or\'us.
\newblock Efficient tensor network simulation of {IBM}'s largest quantum
  processors.
\newblock {\em Phys. Rev. Res.}, 6:013326, Mar 2024.

\bibitem{Kim2023Evidence}
Youngseok Kim, Andrew Eddins, Sajant Anand, Ken~Xuan Wei, Ewout van~den Berg,
  Sami Rosenblatt, Hasan Nayfeh, Yantao Wu, Michael Zaletel, Kristan Temme, and
  Abhinav Kandala.
\newblock Evidence for the utility of quantum computing before fault tolerance.
\newblock {\em Nature}, 618(7965):500--505, 2023.

\bibitem{Fontana2025Classical}
Enrico Fontana, Manuel~S. Rudolph, Ross Duncan, Ivan Rungger, and Cristina
  Cîrstoiu.
\newblock Classical simulations of noisy variational quantum circuits.
\newblock {\em npj Quantum Information}, 11(1):84, 2025.

\bibitem{Begui2025Simulating}
Tomislav Begušić, Kasra Hejazi, and Garnet Kin-Lic Chan.
\newblock Simulating quantum circuit expectation values by {Clifford}
  perturbation theory.
\newblock {\em The Journal of Chemical Physics}, 162(15):154110, 2025.

\bibitem{fb28-wlv2}
Armando Angrisani, Antonio~A. Mele, Manuel~S. Rudolph, M.~Cerezo, and Zo\"e
  Holmes.
\newblock Simulating quantum circuits with arbitrary local noise using {Pauli}
  propagation.
\newblock {\em PRX Quantum}, 7:020313, Apr 2026.

\bibitem{GomezLurbe2026itepp}
Rafael Gómez-Lurbe and Armando Pérez.
\newblock Pauli propagation for imaginary-time evolution.
\newblock {\em New Journal of Physics}, 28(8):084517, 2026.

\bibitem{majumder2026quantumenhancedpaulipropagation}
S.~Majumder, J.~R. Garrison, L.~Luo, B.~Mitchell, M.~Amico, A.~Seif, M.~Tran,
  K.~Sharma, E.~van~den Berg, Z.~Minev, and L.~C.~G. Govia.
\newblock Quantum enhanced {Pauli} propagation, 2026.
\newblock arXiv:2603.14485 [quant-ph].

\bibitem{teng2026leveragingsymmetrymergingpauli}
Yanting Teng, Su~Yeon Chang, Manuel~S. Rudolph, and Zo\"e Holmes.
\newblock Leveraging symmetry merging in {Pauli} propagation, 2026.
\newblock arXiv:2512.12094 [quant-ph].

\bibitem{kroetz2026paulib}
Florian Krötz.
\newblock {PauLIB}: A high-performance library for processing {Pauli} strings,
  2026.
\newblock arXiv:2605.25974 [quant-ph].

\bibitem{SciPostPhysCodeb.54}
Nicolas Loizeau, J.~Clayton Peacock, and Dries Sels.
\newblock Quantum many-body simulations with {PauliStrings.jl}.
\newblock {\em SciPost Phys. Codebases}, 54, 2025.

\bibitem{SciPostPhysCodeb.54-r1.5}
Nicolas Loizeau, J.~Clayton Peacock, and Dries Sels.
\newblock Codebase release 1.5 for {PauliStrings.jl}.
\newblock {\em SciPost Phys. Codebases}, 54-r1.5, 2025.

\bibitem{10.1145/3564246.3585234}
Dorit Aharonov, Xun Gao, Zeph Landau, Yunchao Liu, and Umesh Vazirani.
\newblock A polynomial-time classical algorithm for noisy random circuit
  sampling.
\newblock In {\em Proceedings of the 55th Annual ACM Symposium on Theory of
  Computing}, STOC 2023, pages 945--957, New York, NY, USA, 2023. Association
  for Computing Machinery.

\bibitem{pauli-prop}
Qiskit addons team.
\newblock {Pauli prop}.
\newblock \url{https://github.com/Qiskit/pauli-prop}, 2025.

\bibitem{cuquantum_cupauliprop}
{NVIDIA Corporation}.
\newblock {cuPauliProp}: A high-performance library for {Pauli} propagation
  quantum simulators.
\newblock
  \url{https://docs.nvidia.com/cuda/cuquantum/26.06.0/cupauliprop/index.html},
  2026.
\newblock Version 0.4.0 (cuQuantum SDK 26.06), the version benchmarked here;
  first released in cuQuantum SDK 25.11.

\bibitem{mazumder2026benchmarkingzerosetup}
Arul~Rhik Mazumder, Hovnatan Karapetyan, Mohammed~Zuhair Mullath, Rudy~H.
  Tanin, Hrant Gharibyan, and Hayk Tepanyan.
\newblock Benchmarking zero-setup quantum circuit simulators, 2026.
\newblock arXiv:2607.09882 [quant-ph].

\bibitem{fhc5-8sm6}
Sacha Lerch, Ricard Puig, Manuel~S. Rudolph, Armando Angrisani, Tyson Jones,
  M.~Cerezo, Supanut Thanasilp, and Zo\"e Holmes.
\newblock Efficient quantum-enhanced classical simulation for patches of
  quantum landscapes.
\newblock {\em PRX Quantum}, 7:020359, Jun 2026.

\bibitem{PhysRevLett.131.100803}
Franz~J. Schreiber, Jens Eisert, and Johannes~Jakob Meyer.
\newblock Classical surrogates for quantum learning models.
\newblock {\em Phys. Rev. Lett.}, 131:100803, Sep 2023.

\bibitem{monaco2026symbolicpaulipropagationgradientenabled}
Saverio Monaco, Jamal Slim, Florian Rehm, Dirk Krücker, and Kerstin Borras.
\newblock Symbolic {Pauli} propagation for gradient-enabled pre-training of
  quantum circuits, 2026.
\newblock arXiv:2512.16674 [quant-ph].

\bibitem{8qt9-72ts}
Pablo Bermejo, Paolo Braccia, Manuel~S. Rudolph, Zo\"e Holmes, Lukasz Cincio,
  and M.~Cerezo.
\newblock Quantum convolutional neural networks are effectively classically
  simulable.
\newblock {\em PRX Quantum}, 7:020304, Apr 2026.

\bibitem{Cerezo2025Does}
M.~Cerezo, Martin Larocca, Diego García-Martín, N.~L. Diaz, Paolo Braccia,
  Enrico Fontana, Manuel~S. Rudolph, Pablo Bermejo, Aroosa Ijaz, Supanut
  Thanasilp, Eric~R. Anschuetz, and Zo\"e Holmes.
\newblock Does provable absence of barren plateaus imply classical
  simulability?
\newblock {\em Nature Communications}, 16(1):7907, 2025.

\bibitem{4kyq-n8jb}
Zong-Liang Li and Shi-Xin Zhang.
\newblock Dual role of low-weight {Pauli} propagation: A flawed simulator but a
  powerful initializer for variational quantum algorithms.
\newblock {\em Phys. Rev. Res.}, 8:013266, Mar 2026.

\bibitem{Lin2026utilityscalequantum}
Cheng-Ju Lin, Hrant Gharibyan, and Vincent~P. Su.
\newblock Utility-scale quantum state preparation: classical training using
  {Pauli} path simulation.
\newblock {\em {Quantum}}, 10:2014, March 2026.

\bibitem{gharibyan2025practicalguideusingpauli}
Hrant Gharibyan, Siddharth Hariprakash, Mohammed~Zuhair Mullath, and Vincent~P.
  Su.
\newblock A practical guide to using {Pauli} path simulators for utility-scale
  quantum experiments, 2025.
\newblock arXiv:2507.10771 [quant-ph].

\bibitem{zhang2026stochasticpaulipathsimulatorlargescale}
Kaining Zhang, Xinbiao Wang, Kunsheng Li, Qixin Zhang, Yuxuan Du, Min-Hsiu
  Hsieh, and Dacheng Tao.
\newblock Stochastic {Pauli}-path simulator for large-scale quantum
  optimization, 2026.
\newblock arXiv:2607.17804 [quant-ph].

\bibitem{lin2026backpropagatingpaulipropagation}
Sheng-Hsuan Lin, Etienne Granet, Kevin Hémery, and Henrik Dreyer.
\newblock Backpropagating {Pauli} propagation, 2026.
\newblock arXiv:2607.15184 [quant-ph].

\bibitem{Zhang2023tensorcircuit}
Shi-Xin Zhang, Jonathan Allcock, Zhou-Quan Wan, Shuo Liu, Jiace Sun, Hao Yu,
  Xing-Han Yang, Jiezhong Qiu, Zhaofeng Ye, Yu-Qin Chen, Chee-Kong Lee, Yi-Cong
  Zheng, Shao-Kai Jian, Hong Yao, Chang-Yu Hsieh, and Shengyu Zhang.
\newblock {TensorCircuit}: a quantum software framework for the {NISQ} era.
\newblock {\em {Quantum}}, 7:912, February 2023.

\bibitem{10.1145/3620665.3640366}
Jason Ansel et~al.
\newblock {PyTorch} 2: Faster machine learning through dynamic {Python}
  bytecode transformation and graph compilation.
\newblock In {\em Proceedings of the 29th ACM International Conference on
  Architectural Support for Programming Languages and Operating Systems, Volume
  2}, ASPLOS '24, pages 929--947, New York, NY, USA, 2024. Association for
  Computing Machinery.

\bibitem{j1gg-s6zb}
Victor Martinez, Armando Angrisani, Ekaterina Pankovets, Omar Fawzi, and Daniel
  Stilck Fran\ifmmode~\mbox{\c{c}}\else \c{c}\fi{}a.
\newblock Efficient simulation of parametrized quantum circuits under nonunital
  noise through {Pauli} backpropagation.
\newblock {\em Phys. Rev. Lett.}, 134:250602, Jun 2025.

\bibitem{Mele2026Noise}
Antonio~Anna Mele, Armando Angrisani, Soumik Ghosh, Sumeet Khatri, Jens Eisert,
  Daniel Stilck~França, and Yihui Quek.
\newblock Noise-induced shallow circuits and the absence of barren plateaus.
\newblock {\em Nature Physics}, 22(5):751--756, 2026.

\bibitem{PhysRevA.70.052328}
Scott Aaronson and Daniel Gottesman.
\newblock Improved simulation of stabilizer circuits.
\newblock {\em Phys. Rev. A}, 70:052328, Nov 2004.

\bibitem{9355309}
Trevor Gale, Matei Zaharia, Cliff Young, and Erich Elsen.
\newblock Sparse {GPU} kernels for deep learning.
\newblock In {\em SC20: International Conference for High Performance
  Computing, Networking, Storage and Analysis}, pages 1--14, 2020.

\bibitem{paulipropagation_jl_repo}
Manuel~S. Rudolph et~al.
\newblock {PauliPropagation.jl}.
\newblock \url{https://github.com/SparqleSim/PauliPropagation.jl}, 2026.
\newblock Version 0.8.1; version 0.7.0 (February 2026) added the experimental
  CUDA extension and version 0.8.0 (August 2026) \texttt{rewindgradient}.

\bibitem{jones2020efficientcalculationgradientsclassical}
Tyson Jones and Julien Gacon.
\newblock Efficient calculation of gradients in classical simulations of
  variational quantum algorithms, 2020.
\newblock arXiv:2009.02823 [quant-ph].

\bibitem{PhysRevA.103.032430}
Maria Schuld, Ryan Sweke, and Johannes~Jakob Meyer.
\newblock Effect of data encoding on the expressive power of variational
  quantum-machine-learning models.
\newblock {\em Phys. Rev. A}, 103:032430, Mar 2021.

\bibitem{PerezSalinas2020datareuploading}
Adri{\'{a}}n P{\'{e}}rez-Salinas, Alba Cervera-Lierta, Elies Gil-Fuster, and
  Jos{\'{e}}~I. Latorre.
\newblock Data re-uploading for a universal quantum classifier.
\newblock {\em {Quantum}}, 4:226, February 2020.

\bibitem{herrerogonzalez2025bornultimatumconditionsclassical}
Mario Herrero-Gonzalez, Brian Coyle, Kieran McDowall, Ross Grassie, Sjoerd
  Beentjes, Ava Khamseh, and Elham Kashefi.
\newblock The {Born} ultimatum: Conditions for classical surrogation of quantum
  generative models with correlators, 2025.
\newblock arXiv:2511.01845 [quant-ph].

\bibitem{cirstoiu2024fourieranalysisframeworkapproximate}
Cristina Cîrstoiu.
\newblock A {Fourier} analysis framework for approximate classical simulations
  of quantum circuits, 2024.
\newblock arXiv:2410.13856 [quant-ph].

\bibitem{PhysRevLett.115.070501}
Hakop Pashayan, Joel~J. Wallman, and Stephen~D. Bartlett.
\newblock Estimating outcome probabilities of quantum circuits using
  quasiprobabilities.
\newblock {\em Phys. Rev. Lett.}, 115:070501, Aug 2015.

\bibitem{bergholm2022pennylaneautomaticdifferentiationhybrid}
Ville Bergholm et~al.
\newblock {PennyLane}: Automatic differentiation of hybrid quantum-classical
  computations, 2022.
\newblock arXiv:1811.04968 [quant-ph].

\bibitem{Herrman2022}
Rebekah Herrman, Phillip~C. Lotshaw, James Ostrowski, Travis~S. Humble, and
  George Siopsis.
\newblock Multi-angle quantum approximate optimization algorithm.
\newblock {\em Scientific Reports}, 12(1):6781, Apr 2022.

\bibitem{asadi2024hybridquantumprogrammingpennylane}
Ali Asadi, Amintor Dusko, Chae-Yeun Park, Vincent Michaud-Rioux, Isidor Schoch,
  Shuli Shu, Trevor Vincent, and Lee~James O'Riordan.
\newblock Hybrid quantum programming with {PennyLane} {Lightning} on {HPC}
  platforms, 2024.
\newblock arXiv:2403.02512 [quant-ph].

\bibitem{PhysRevA.99.032331}
Maria Schuld, Ville Bergholm, Christian Gogolin, Josh Izaac, and Nathan
  Killoran.
\newblock Evaluating analytic gradients on quantum hardware.
\newblock {\em Phys. Rev. A}, 99:032331, Mar 2019.

\bibitem{Hirviniemi:2026dzs}
Olli Hirviniemi, Afrad Basheer, and Thomas Cope.
\newblock Random quantum circuits as seeds for continuous generative models,
  2026.
\newblock arXiv:2602.10049 [quant-ph].

\bibitem{Peruzzo2014}
Alberto Peruzzo, Jarrod McClean, Peter Shadbolt, Man-Hong Yung, Xiao-Qi Zhou,
  Peter~J. Love, Al{\'a}n Aspuru-Guzik, and Jeremy~L. O'Brien.
\newblock A variational eigenvalue solver on a photonic quantum processor.
\newblock {\em Nature Communications}, 5(1):4213, Jul 2014.

\bibitem{farhi2014quantumapproximateoptimizationalgorithm}
Edward Farhi, Jeffrey Goldstone, and Sam Gutmann.
\newblock A quantum approximate optimization algorithm, 2014.
\newblock arXiv:1411.4028 [quant-ph].

\bibitem{kim2026safemaqaoasurrogateassistedfinetuning}
Hyunwoo Kim and Youngseok Lee.
\newblock {SAFE} ma-{QAOA}: Surrogate-assisted and fine-tuning enhanced
  multi-angle {QAOA} with parameter distillation, 2026.
\newblock arXiv:2605.23377 [quant-ph].

\bibitem{GonzalezGarcia2025paulipath}
Guillermo Gonz{\'{a}}lez-Garc{\'{i}}a, J.~Ignacio Cirac, and Rahul Trivedi.
\newblock Pauli path simulations of noisy quantum circuits beyond average case.
\newblock {\em {Quantum}}, 9:1730, May 2025.

\bibitem{lh6x-7rc3}
Armando Angrisani, Alexander Schmidhuber, Manuel~S. Rudolph, M.~Cerezo, Zo\"e
  Holmes, and Hsin-Yuan Huang.
\newblock Classically estimating observables of noiseless quantum circuits.
\newblock {\em Phys. Rev. Lett.}, 135:170602, Oct 2025.

\end{thebibliography}
}

\clearpage
\appendix
\crefalias{section}{appendix}
\crefalias{subsection}{appendix}

\section{Core Data Structures and Internal Routines}
\label{app:internals}

The objects below are what the calls of \Cref{sec:compile-once,sec:api} compile to and run on. They are importable, so that custom evaluation or instrumentation code can be written without modifying the engine; the supported interface is the methods of \texttt{CompiledProgram} and of the sampler, and the compile profiler at the end of this list. Beyond the names of \Cref{sec:compile-once,sec:api}, the top-level namespace also exports \path{pennylane_apply_circuit} and \path{pennylane_observable}, the two functions the reference helpers are built from, so that the same cross-check can be assembled on another PennyLane device. \Cref{lst:objects} shows how the objects fit together.

\begin{lstlisting}[language={},float=tb,columns=fixed,basewidth=0.55em,caption={The compiled objects and what they hold.},label={lst:objects}]
Circuit.compile(...)  [calls compile_program(...)]
  -> CompiledProgram
       |-- circuit, observables                     gate list; observables as PauliSum
       |-- psum_union : TensorOperatorProgram       the compiled program
       |     |-- x_mask, z_mask, coeff_init         output rows, coefficients c_init
       |     +-- steps : [TensorSparseStep]         one per gate
       |           |-- mat_const, mat_cos, mat_sin  step matrices
       |           +-- param_idx | embedding_idx    the angle the step reads
       |-- basis : UnionBasis                       surviving Pauli strings
       |-- preset : ExecutionPreset
       |-- const_thetas                             fixed angles from rot(angle=...)
       |-- compile_stats                            profiler output
       |-- expvals(), expval()                      run through TensorSparseEvaluator
       +-- expvals_reference(), clear_cache()       PennyLane reference; cache release
build_quasi_sampler(...)
  -> TensorSparseSampler : compute_moments(), quasi_prob(), quasi_prob_from_moments()
\end{lstlisting}

\begin{itemize}[leftmargin=1.2em,itemsep=0.5em]
    \item \texttt{TensorOperatorProgram} and \texttt{TensorSparseStep}: the compiled program of \Cref{sec:compile-once} and its sparse steps. A \texttt{CompiledProgram} holds the program as \texttt{psum\_union}, next to the shrunk \texttt{UnionBasis} (\texttt{basis}) and \texttt{compile\_stats}. The program stores its output rows (\texttt{x\_mask}, \texttt{z\_mask}), the coefficient vector $\mathbf{c}_{\mathrm{init}}$ (\texttt{coeff\_init}) of the observable on the union basis described under \path{build_union_observable} below, and the steps. Each step stores $M_{\mathrm{const}}$, $M_{\cos}$, and $M_{\sin}$ as \texttt{mat\_const}, \texttt{mat\_cos}, and \texttt{mat\_sin}, together with the angle it reads (\texttt{param\_idx}, or \texttt{embedding\_idx} for a rotation driven by an embedding input). A Pauli rotation step keeps its unchanged branches as an index map (\texttt{unchanged\_cols}, with \texttt{cos\_pos} marking the rows that carry $\cos\theta$) and its sine branches as a matrix; Clifford and noise steps use \texttt{mat\_const} alone.
    \item \texttt{TensorSparseSampler}: the quasi-probability sampler that \path{build_quasi_sampler} returns, with the \texttt{compute\_moments}, \texttt{quasi\_prob}, and \path{quasi_prob_from_moments} methods of \Cref{sec:api}.
    \item \texttt{TensorSparseEvaluator}: the class that runs a \texttt{TensorOperatorProgram}. Its \path{evaluate_adjoint} applies the transposed step matrices, in sequence, to the survival mask $\mathbf{m}_{L+1}$ for a given parameter vector, caching those matrices between calls or, for a step matrix stored off the compute device with more than \texttt{chunk\_size} rows, streaming it in blocks of that size without caching. A compiled program instantiates it internally; construct it directly only for custom evaluation schedules.
    \item \path{build_union_observable(observables)} and \texttt{UnionBasis}: the first compile stage. It builds the shared observable on the union basis over all listed observables and returns it together with a \texttt{UnionBasis}, the ordered Pauli string list and its index. The coefficient the returned observable carries for each string is the largest $\lvert c\rvert$ that string has in any listed observable ($1$ when that is zero); it serves only the coefficient truncation that \path{build_min_abs} enables, while $V_0$ is built from the observables' own coefficients. The second and third stages are \path{propagate_tensor_program} and \path{zero_filter_tensor_backprop_with_keep_mask}, whose keep mask \path{shrink_union_basis} applies to the \texttt{UnionBasis}.
    \item \path{adjoint_weights_on_zero(...)} and \path{coeff_matrix_from_observables(...)}: the readout in two stages behind \texttt{expvals}. \path{adjoint_weights_on_zero} produces $\mathbf{w}(\bm\theta)$ by an adjoint pass, \path{coeff_matrix_from_observables} assembles the observable coefficients into the dense matrix $V_0$, and the expectation values follow from the single product $\mathbf{E}=\mathbf{w}^{\top}V_0$. Call these directly to share one adjoint pass across any observables whose Pauli strings lie in the compiled basis.
    \item \texttt{CompiledProgram.compile\_stats} and the \path{PPS_COMPILE_PROFILE} environment variable: the compile profiler, and the supported way to obtain the compilation times and Pauli term counts of \Cref{sec:trunc-cost} and \Cref{sec:benchmarks} without instrumenting the engine. With \path{PPS_COMPILE_PROFILE=1} set, \path{compile_program} times each stage and records on the returned program's \texttt{compile\_stats} a per-step breakdown of propagation and of zero-filtering, plus a \texttt{summary} entry with the stage totals and the propagated and surviving term counts.
\end{itemize}

\end{document}